\documentclass[12pt,letterpaper]{article}

\usepackage{graphicx,array}
\usepackage{url}
\usepackage{color}
\usepackage{latexsym}
\usepackage{amsthm}
\usepackage{amsmath}
\usepackage{amssymb}
\usepackage{amsfonts}
\usepackage[numbers,sort&compress]{natbib}
\usepackage{bm}
\usepackage{slashed}
\usepackage{mathrsfs}
\usepackage{enumerate}
\usepackage{tikz}
\usepackage{siunitx}
\usepackage{mdframed}
\usepackage{setspace}  
\usepackage{esvect}
\usepackage{esint}
\usepackage{subcaption}
\usepackage[normalem]{ulem}
\usepackage{setspace}

\usepackage{tcolorbox}%

\usepackage{hyperref} 

\hypersetup{
    colorlinks=true,       
    linkcolor=red,          
    citecolor=blue,        
    filecolor=magenta,      
    urlcolor=blue           
}

\usepackage[all]{hypcap} 

\usepackage{natbib}
\newcommand{\nc}{\newcommand}

\nc{\beq}{\begin{equation}}  
\nc{\eeq}{\end{equation}}  
\nc{\beqa}{\begin{eqnarray}}  
\nc{\eeqa}{\end{eqnarray}}  
\nc{\bit}{\begin{itemize}}  
\nc{\eit}{\end{itemize}}

\usepackage{floatrow}
\newfloatcommand{capbtabbox}{table}[][\FBwidth]

\usepackage{blindtext}

\title{ 
\rightline{\tiny KCL-PH-TH/2023-44}
\setlength{\parskip}{2ex}
{\bf A Cosmic Window on the Dark Axion Portal}
\author{\large Heejoung Hong$^{\, a}$, Ui Min$^{\, a}$, Minho Son$^{\, a}$, and Tevong You$^{\, b}$}
\date{\small \it 
$^a$Department of Physics, Korea Advanced Institute of Science and Technology, \\
291 Daehak-ro, Yuseong-gu, Daejeon 34141, Republic of Korea\\
$^b$Theoretical Particle Physics and Cosmology Group, Department of Physics, \\ King’s College London, London, WC2R 2LS, UK
}
}

\begin{document}

\maketitle

\setlength{\parskip}{0.2ex}

\begin{abstract}	
Axions and dark photons are common in many extensions of the Standard Model. The dark axion portal---an axion coupling to the dark photon and photon---can significantly modify their phenomenology. We study the cosmological constraints on the dark axion portal from Cosmic Microwave Background (CMB) bounds on the energy density of dark radiation, $\Delta N_\text{eff}$. By computing the axion-photon-dark photon collision terms and solving the Boltzmann equations including their effects, we find that light axions are generally more constrained by $\Delta N_\text{eff}$ than from supernova cooling or collider experiments. However, with dark photons at the MeV scale, a window of parameter space is opened up above the supernova limits and below the experimental exclusion, allowing for axion decay constants as low as $f_a \sim 10^4$ GeV. This region also modifies indirectly the neutrino energy density, thus relaxing the cosmological upper bound on the sum of neutrino masses. Future CMB measurements could detect a signal or close this open window on the dark axion portal.
\end{abstract}

\thispagestyle{empty}  
\newpage  
  
\setcounter{page}{1}

\begingroup
\hypersetup{linkcolor=black,linktocpage}
\tableofcontents
\endgroup




\section{Introduction}
\label{sec:intro}

Standard Model (SM) extensions featuring dark sectors arise in a number of theories beyond the Standard Model (BSM). The evidence for dark matter and neutrino masses also implies the existence of a dark sector. Such a dark sector may consist of a broad spectrum of particles, in the same way that our visible sector is described by a highly non-minimal SM. This motivates studying the potential phenomenological effects of an extended dark sector for a variety of non-minimal scenarios, to anticipate ways in which they may manifest themselves in observations and experiments. 

Astrophysical and cosmological observations place some of the strongest constraints on light dark sectors up to masses around the MeV scale and at weaker couplings than accessible to direct searches in experiment. Supernovae emissions in particular are sensitive to additional weakly coupled new degrees of freedom that provide an extra source of energy loss beyond neutrinos~\cite{Raffelt:1996wa}. Since neutrinos have been detected from the explosion of supernova SN1987A~\cite{Kamiokande-II:1987idp,Bionta:1987qt,Alekseev:1987ej}, supernova cooling has been used to constrain a plethora of BSM scenarios. However, there is an upper bound on supernovae exclusion limits when the couplings become strong enough to no longer enable efficient cooling, and excluding new physics in this process is subject to uncertainties in supernova modelling. Cosmological bounds on dark radiation from the Cosmic Microwave Background (CMB) are more robust. They have previously been applied to dark photons and axions, for example, in Refs.~\cite{Ibe:2019gpv, Baumann:2016wac}. We investigate here their implications for a non-minimal scenario consisting of two light dark degrees of freedom. In the dynamics of cosmological evolution, described by the Boltzmann equation, there may be a non-trivial interplay between the energy density in visible radiation, dark radiation, and neutrinos that lead to unconstrained regions of parameter space which would naively have been excluded. We point out such a possibility here in a BSM extension involving both an axion and a dark photon.

Axions and dark photons are two of the most widely studied dark sector candidates (see for example Refs.~\cite{Fabbrichesi:2020wbt, Marsh:2015xka} for a review). Either of them could constitute dark matter, but in any case they could still be part of the dark sector. The existence of both in a non-minimal dark sector implies an axion-photon-dark photon interaction known as the dark axion portal; in an Effective Field Theory (EFT) description, all terms allowed by symmetries are a priori present. The Lagrangian term for the dark axion portal interaction is given by 
\begin{equation}\label{eq:portal:Lag}
   \mathcal{L} \supset \frac{a}{2 f_a} F_{\mu\nu}\tilde{F'}^{\mu\nu}~,
\end{equation}
where $a$ is an axion-like particle with mass $m_a$, $F_{\mu\nu}$ is the field strength of the massless photon $A_\mu$, $\tilde{F'}_{\mu\nu}$ is the dual field strength of the dark photon ${A'}_\mu$ of mass $m_{\gamma'}$, and $f_a$ is the axion decay constant. The relative strengths of the various EFT interactions depend on the ultra-violet (UV) completion. Here we take the dark axion portal to be the dominant interaction between visible and dark sectors. For example, one may assume a charge conjugate symmetry in the dark sector under which the axion is odd to justify the axion-photon-dark photon coupling as the leading interaction while preventing the kinetic mixing of the dark photon to the SM. The dark axion portal of Eq.~(\ref{eq:portal:Lag}) has previously been explored in various contexts~\cite{Kaneta:2016wvf,Kaneta:2017wfh,Pospelov:2018kdh,Choi:2018mvk,Kalashev:2018bra,Biswas:2019lcp,Choi:2019jwx,Hook:2019hdk,Deniverville:2020rbv,Arias:2020tzl,Hook:2021ous,Domcke:2021yuz,Gutierrez:2021gol,Carenza:2023qxh,Lane:2023eno,Jodlowski:2023sbi, Jodlowski:2023yne, Hook:2023smg}. It may also play a role in a mechanism for cosmological relaxation of the Higgs mass~\cite{Domcke:2021yuz}.

In this work, we compute the axion-photon-dark photon collision terms when solving the Boltzmann equations to determine $\Delta N_\text{eff}$, the additional relativistic degrees of freedom in light species other than neutrinos. The effect of the dark axion portal on the allowed dark radiation contribution from a light axion is drastically altered by the presence of a dark photon when the dark photon's mass is around the MeV scale. We find that in the excluded region $f_a \lesssim 10^7$ GeV, there is an open window of parameter space when $m_{\gamma^\prime}$ is between 1 and 10 MeV where $f_a$ can be as large as $10^4$ GeV. This is due to the dark photon decay after the time of Big Bang Nucleosynthesis (BBN) but before the CMB recombination era, which proceeds through the dark axion portal interaction to a massless photon and a light axion. The contribution to the visible radiation is sufficient to indirectly alter the neutrino energy density away from its value in the SM. The axion dark radiation contribution to $\Delta N_\text{eff}$ can then be larger while remaining in agreement with CMB constraints from Planck data.   

This island of parameter space is only partly constrained by supernova cooling bounds. For stronger couplings, or equivalently smaller axion decay constants $f_a \lesssim 10^5$ GeV, supernova trapping prevents extending the supernova exclusion reach. Constraints from collider experiments, in particular CHARM~\cite{CHARM:1985anb}, only enter below $f_a \lesssim 10^4$ GeV. The experimental sensitivity can be improved by the proposed SHiP experiment~\cite{SHiP:2015vad,Alekhin:2015byh}, though not enough to cover the mass-coupling range necessary to close the island region. There is therefore an open window of parameter space for the dark axion portal that is currently unconstrained but could be comprehensively probed by future CMB measurements such as CMB-S4~\cite{CMB-S4:2016ple}. 

The alteration of the neutrino energy density in this window also relaxes the cosmological upper bound on the sum of neutrino masses. In particular, for certain cases this may reopen the possibility of inverted ordering which is currently excluded in the standard scenario by Planck constraints~\cite{DiValentino:2021hoh}. A complementary probe is provided by the next generation of neutrino experiments. They are projected to reach the sensitivity necessary to cover the entire inverted ordering parameter space. Should inverted ordering by preferred, the dark axion portal has the potential to alleviate a small tension between the cosmological bound and terrestrial determinations of neutrino masses.

This paper is organised as follows: in Section~\ref{sec:Boltzmann} we review the Boltzmann equations that are used to describe the cosmological evolution of the energy densities of light relativistic species. In Section~\ref{sec:Neff} we describe the results of our numerical simulation to determine $\Delta N_\text{eff}$ for the case of a massive dark photon and light axion, and vice versa. Finally, in Section~\ref{sec:neutrino}, we show how the modification of the neutrino energy density in the open window of parameter space can relax the cosmological bound on the sum of neutrino masses. 
In Section~\ref{sec:BBN}, we discuss the cosmological implications for Big Bang Nucleosynthesis (BBN).
We conclude in Section~\ref{sec:conclusion}. Expressions for the collision terms are collected in the Appendix.

%
\section{Boltzmann equations}
\label{sec:Boltzmann}
The cosmological evolution of the energy densities of the degrees of freedom present in the thermal bath of the early universe is well described by the Boltzmann equations. The energy density of relatistivic species is parametrised in terms of $N_\text{eff}$, the effective number of neutrino species. We solve the Boltzmann equations numerically starting with a temperature around the muon mass when it is a good assumption that the universe was in a plasma state consisting of photons, electrons, positrons, and neutrinos in the SM. We include extra relativistic BSM species that can alter the cosmological evolution of the energy density and lead to an interesting cosmological signature

\begin{figure}[t]
\begin{center}
\includegraphics[width=0.33\textwidth]{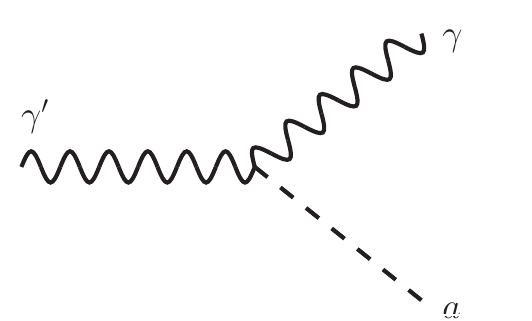}
\caption{\small Feynman diagram for the dark axion portal interaction involving a photon $\gamma$, dark photon $\gamma^\prime$, and axion $a$, contributing to the three-point collision term, $C^{\gamma',\, a}_{\gamma' \leftrightarrow a \gamma}$.}
\label{fig:3pt:diagrams}
\end{center}
\end{figure}

\begin{figure}[t]
\begin{center}
\includegraphics[width=0.33\textwidth]{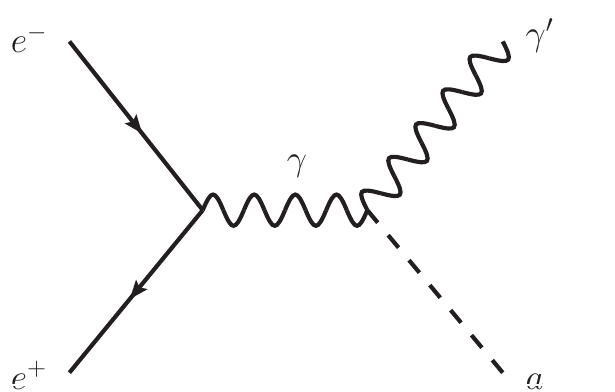}
\includegraphics[width=0.33\textwidth]{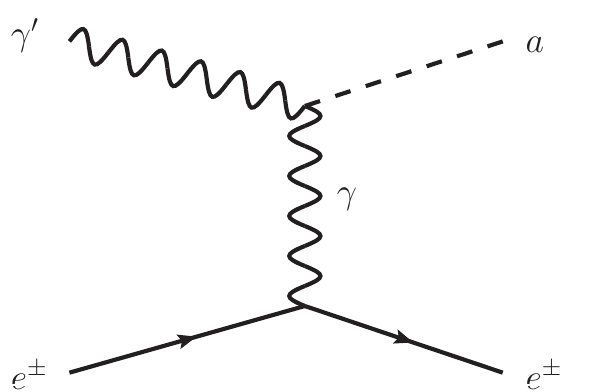}
\caption{\small Feynman diagrams involving scattering between electrons and positrons $e^\pm$, the dark photon $\gamma^\prime$, and axion $a$ that contribute to four-point collision terms $C^{\gamma',\, a}_{\gamma' a \leftrightarrow e^+ e^-}$ and $C^{\gamma',\, a}_{e^\pm \gamma' \leftrightarrow e^\pm a}$. }
\label{fig:4pt:diagrams}
\end{center}
\end{figure}

The Boltzmann equation takes the form, 
\begin{equation}\label{eq:Boltzmann}
  \frac{d\rho_i}{dt} + 3 H \left ( \rho_i + p_i \right ) = - \sum_a C^i_a~,
\end{equation}
where $\rho_i$ is the energy density and $C^i_a$ is the collision term of a given species $i$ for the process $a$.
The Boltzmann equations for the SM are taken from Refs.~\cite{Escudero:2018mvt, Hannestad:1995rs, Kawasaki:2000en, Kawasaki:1999na}. 
The additional Boltzmann equations for the dark photon and axion and the modification of the SM part in the dark axion portal model are given by
\begin{equation}
\begin{split}
  \frac{d\rho_{\gamma'}}{dt} + 3 H \left ( \rho_{\gamma'} + p_{\gamma'} \right ) &= - C^{\gamma'}_{\gamma' \leftrightarrow a \gamma}
  - C^{\gamma'}_{\gamma' a \leftrightarrow e^+ e^-} - 2\, C^{\gamma'}_{e^\pm \gamma' \leftrightarrow e^\pm a}~,
  \\[5pt]
    \frac{d\rho_{a}}{dt} + 3 H \left (\rho_{a} + p_{a} \right ) &= C^{a}_{\gamma' \leftrightarrow a \gamma}
  - C^{a}_{\gamma' a \leftrightarrow e^+ e^-} + 2\, C^{a}_{e^\pm \gamma' \leftrightarrow e^\pm a}~,
    \\[5pt]
     \frac{d\rho_{\gamma e}}{dt} + 3 H \left (\rho_{\gamma e} + p_{\gamma e} \right ) &= 
     - C_{e\leftrightarrow \nu_e}  -2\, C_{e\leftrightarrow \nu_\mu, \nu_\tau}
     + C^{\gamma}_{\gamma' \leftrightarrow a \gamma}
  + C^{e}_{\gamma' a \leftrightarrow e^+ e^-} -2\, C^{e}_{e^\pm \gamma' \leftrightarrow e^\pm a}~, 
\end{split}
\end{equation}
where the explicit forms of collision terms are given in Appendix~\ref{app:collision}. The collision terms involve five temperatures, which are different a priori, namely, $T_{\nu_e}$,  $T_{\nu_\mu} = T_{\nu_\tau}$ for neutrinos,  $T_{e\gamma} = T_\gamma = T_e$ for the electron-photon (see e.g. Ref.~\cite{Escudero:2018mvt} for related discussion), $T_a$ for the axion, and $T_{\gamma'}$ for the dark photon. Since we start with an initial temperature around the muon mass, only electrons were included in the Boltzmann equations and the electron neutrinos interact differently from muon and tau neutrinos. 
The 3-point collision term gets a contribution from the process depicted in Fig.~\ref{fig:3pt:diagrams}. The diagrams for the leading 4-point collision terms in the coupling strength proportional to $1/f_a$ are shown in Fig.~\ref{fig:4pt:diagrams}.

In numerically solving the Boltzmann equations, we do not include the chemical potential (which is negligible for the electron and neutrino and vanishes for the axion, photon and dark photon) and make some assumptions regarding the statistics of the particles in the collision terms. For the process of $1 \leftrightarrow 2+3$ with the particle 1 being the heaviest, we take into account the quantum statistics only for particle 1 and assume the Maxwell-Boltzmann distributions for particles 2 and 3, such that the integrand of the collision term is proportional to $\big [ f_1  - f_2 f_3 (1\pm f_1)  \big ]$, where $f_i$ is the distribution function. For the process of $1 + 2 \leftrightarrow 3 + 4$, we assume the Maxwell-Boltzmann distributions for all particles which is equivalent to setting $\big [ f_1 f_2  - f_3 f_4  \big ]$ in the integrand of the collision terms. The above assumptions are valid at least for the heavy dark photon (or heavy axion) scenario in the mass range above 1 MeV. The 4-point collision terms can be dominant only at a high temperature where all the particles are well approximated as relativistic following the Maxwell-Boltzmann distributions. At a lower temperature, the 3-point collision term becomes dominant and the lighter particle and photon can be approximated with the Maxwell-Boltzmann distribution, whereas the quantum statistics of the heavier particle is still taken into account.

New degrees of freedom in the thermal bath during Big Bang Nucleosynthesis (BBN) can modify the observed primordial abundances of helium and deuterium, for instance, through the Boltzmann equations of protons and neutrons~\cite{Burns:2023sgx}:
\begin{equation}\label{eq:BBN:pn}
\begin{split}
  \frac{dX_n}{dt} = \Gamma_{p \rightarrow n} X_p - \Gamma_{n\rightarrow p} X_n~, \quad
  \frac{dX_p}{dt} = \Gamma_{n \rightarrow p} X_n - \Gamma_{p\rightarrow n} X_p~,
\end{split}
\end{equation}
where $X_i \equiv n_i/n_B$ and $\Gamma_{n\rightarrow p} = \Gamma(ne^+ \rightarrow p \bar{\nu}) + \Gamma(n\bar{\nu} \rightarrow pe^-) + \Gamma(n\rightarrow p e \bar{\nu})$.
Note that the Hubble parameter, possibly modified due to new degrees of freedom, does not appear in Eq.~\ref{eq:BBN:pn}. 
Since the overall Boltzmann suppression factors on both sides of Eq.~\ref{eq:BBN:pn} cancel out due to the BBN temperature being much smaller than the nucleon mass scale, they prevent terms in Eq.~\ref{eq:BBN:pn} from influencing those in Eq.~\ref{eq:Boltzmann} from which the neutrino and photon temperatures are dominantly determined. However, the modified temperatures of neutrinos and photons feed into the Boltzmann equations in Eq.~\ref{eq:BBN:pn} and may alter the cosmological evolution of primordial abundances. This will be discussed in detail in Section~\ref{sec:BBN}.
While the collision term $\Gamma(\gamma' +p \leftrightarrow a + p)$, mediated by a $t$-channel photon, is allowed, it will not alter the evolution of the proton abundance. The collision term will also not affect the evolution in Eq.~\ref{eq:Boltzmann} around the BBN epoch, due to the Boltzmann suppression.

A naive dimensional analysis estimate for the BBN bound was placed in Ref.~\cite{Hook:2021ous}. They find $f_a^{-1} \lesssim 6.5 \times 10^{-7}$ GeV$^{-1}$ for massless axions and $m_{\gamma'} < 1$ MeV. The constraint is based on requiring $\Gamma \leq H$, where $\Gamma$ is the thermally averaged interaction rate around $T = T_\text{BBN}$ for $e^+e^-\rightarrow a\gamma'$. This requirement effectively puts a bound on $\Delta N_\text{eff}$ as it prevents axions and dark photons from being thermalised such that axions and dark photons do not contribute to the number of effective neutrinos. However, as $\Gamma \propto T^3$ and $H\propto T^2$, the BSM species can be thermalised at higher temperatures even if they do not at BBN. In this case the cosmological constraints from $N_\text{eff}$ at the time of CMB are more relevant than BBN. In the next Section we will focus on computing the energy density of relativistic species at CMB temperatures.

%
\section{Cosmological constraint from $N_\text{eff}$}
\label{sec:Neff}

Using the Boltzmann equations described above to solve numerically for the energy density of radiation, we may now place cosmological constraints on any extra contributions from dark sectors. The effective number of neutrino species is defined as
\begin{equation}
\begin{split}
   N_\text{eff} &\equiv \frac{8}{7}\left ( \frac{11}{4} \right )^{4/3} \left ( \frac{\rho_\text{rad} - \rho_\gamma}{\rho_\gamma} \right )~.
\end{split}
\end{equation}
We adopt the constraint $N_\text{eff} = 2.99^{+0.34}_{-0.33}$ at 95\% CL as determined by Planck 2018 combined with polarisation, lensing, and Baryon Acoustic Oscillation (BAO) data~\cite{Planck:2018vyg}. The value in the SM was recently updated to be 3.043~\cite{Cielo:2023bqp,Bennett:2019ewm,Mangano:2001iu,Mangano:2005cc,deSalas:2016ztq,Akita:2020szl,Froustey:2020mcq,Bennett:2020zkv}.

We take the axion mass to be lighter than about $\mu$eV to avoid exceeding the current dark matter relic abundance, while the dark photon mass is fixed at around the MeV scale. The opposite scenario with a light dark photon and MeV-scale axion will be considered at the end of this Section. We will not pursue the scenario where both the axion and dark photon are stable and massive enough to behave like non-relativistic particles, which would typically require an additional decay channel to avoid having too large a dark matter relic abundance.

The MeV-scale dark photon that is produced in the thermal bath when the temperature was sufficiently high will decay away at some point between BBN and CMB times. The decay proceeds through the dark axion portal to photons and axions, and thus contributes to both visible and dark radiation. In addition to the direct dark radiation contribution to $N_\text{eff}$ from the axion, the modified photon energy density means that there is also an indirect modification to the subsequent evolution of the neutrino energy density. The net effect of this joint change is to allow for a wider range of axion and neutrino energy densities at CMB recombination than would otherwise be compatible with the observed $N_\text{eff}$. 

\begin{figure}[tph]
\begin{center}
\includegraphics[width=0.40\textwidth]{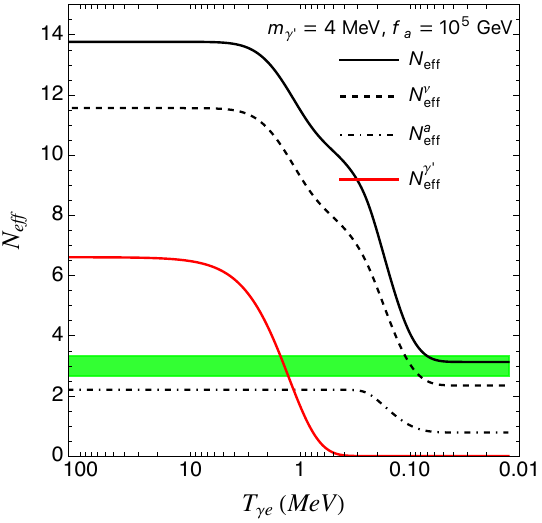}
\includegraphics[width=0.4052\textwidth]{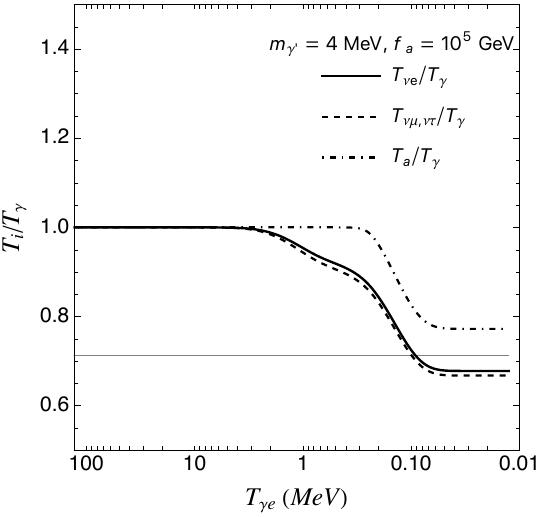}\\
\includegraphics[width=0.40\textwidth]{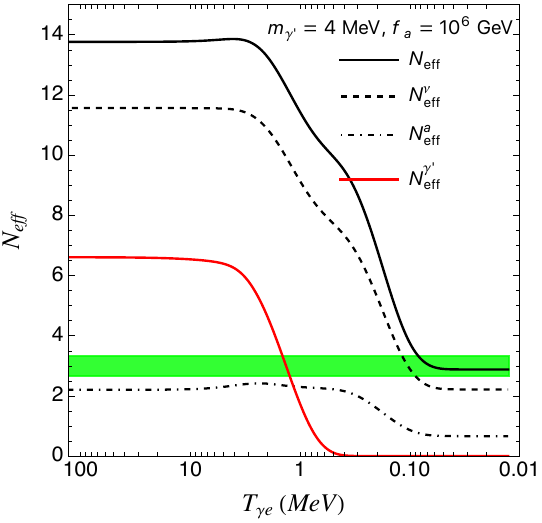}
\includegraphics[width=0.4052\textwidth]{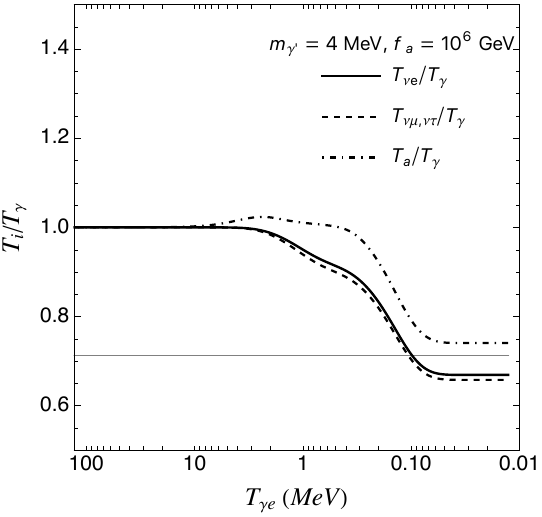}\\
\includegraphics[width=0.40\textwidth]{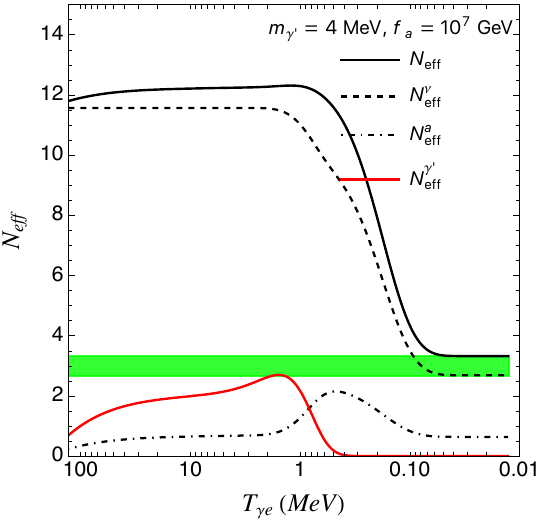}
\includegraphics[width=0.4052\textwidth]{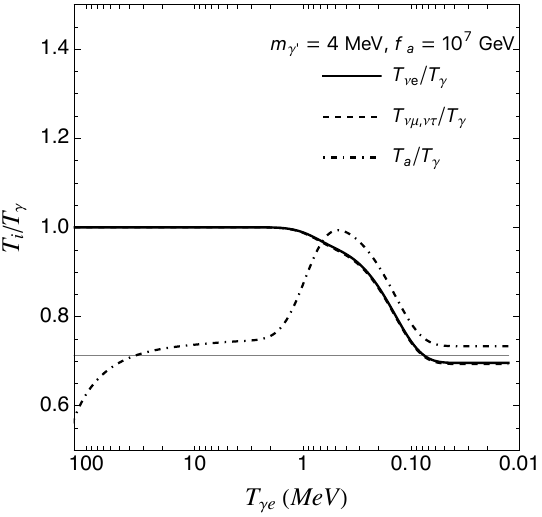}
\caption{\small Left:  the evolutions of $N_\text{eff} \equiv N_\text{eff}^\nu + N_\text{eff}^a$ and individual contributions for $m_{\gamma'} = 4$ MeV and $f_a = 10^{5}, 10^{6}, 10^{7}$ GeV. The green band corresponds to the allowed region, $N_\text{eff} = 2.99^{+0.34}_{-0.33}$ from Planck 2018~\cite{Planck:2018vyg}. Right: the evolution of the $i$-species-to-photon temperature ratios for the same benchmark scenarios. The grey horizontal line corresponds to $T_\nu/T_\gamma = (4/11)^{1/3}$.}
\label{fig:Neff:evol:subset:vert}
\end{center}
\end{figure}

This effect can be seen qualitatively by expressing $N_\text{eff}$ in terms of the neutrino, photon, and axion temperatures, assuming the axions to be relativistic when they decoupled, as 
\begin{equation}\label{eq:Neff:T:darkphoton}
  N_\text{eff} 
  =  \left ( \frac{11}{4} \right )^{4/3}   \left [ 3 \left ( \frac{T_\nu}{T_\gamma} \right )^4 + \frac{8}{7}\frac{g_a}{g_\gamma} \left ( \frac{T_a}{T_\gamma} \right )^4 \right ]~,
\end{equation}
where the first term takes into account a different neutrino-to-photon temperature ratio from the typical value $\left ( 4/11 \right )^{1/3}$ in the SM depending on the dark photon mass and the decoupling epoch of the light axion, and the second term represents the axion dark radiation contribution with $g_a$ and $g_\gamma$ parametrising the number of degrees of freedom of the axion and photon respectively.
%
\begin{figure}[t]
\begin{center}
\includegraphics[width=0.443\textwidth]{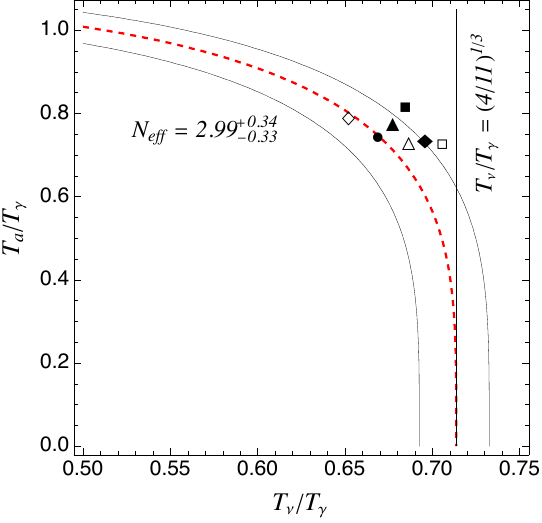}\quad
\includegraphics[width=0.45\textwidth]{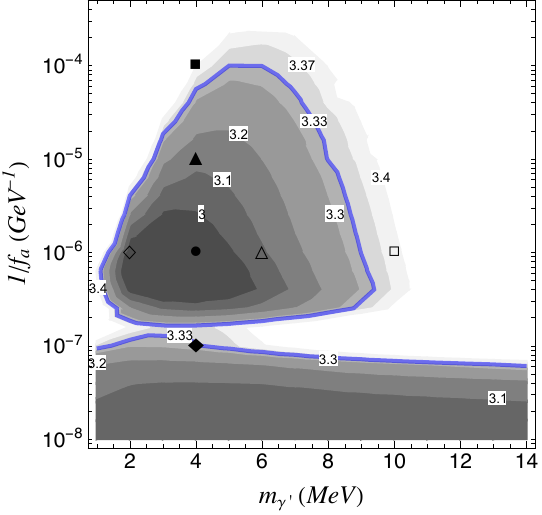}
\caption{\small Left: Allowed $N_\text{eff}$ band for the neutrino-to-photon and axion-to-photon temperature ratios assuming that the decoupling of a particle is instantaneous for the heavy dark photon and light axion scenario. Right: Exclusion plot in the plane $(m_{\gamma'},\, 1/f_a)$ showing the contours of $N_\text{eff}$ values. The dark grey region inside the island bounded by the blue solid line and below the horizontal solid blue line with $N_\text{eff} < 3.33$ are allowed.}
\label{fig:temp}
\end{center}
\end{figure}
%
The neutrino and axion temperatures consistent with observation is illustrated in the left panel of Fig.~\ref{fig:temp}. We see that a modified neutrino temperature allows higher values of the axion temperature that would otherwise be disfavoured for a SM-like neutrino temperature.  The points on the plot correspond to the points shown on the right panel of Fig.~\ref{fig:temp} in the plane of $m_{\gamma'}$ vs $1/f_a$. In that figure, the grey region within and below the blue solid line contours denote the allowed $N_\text{eff} < 3.33$ parameter space as computed from our numerical simulation. For an axion in thermal equilibrium, the axion temperature $T_a$ will be the same as the photon radiation temperature $T_\gamma$. A smaller axion temperature can only be achieved by decoupling the axion from the thermal bath earlier, which happens for sufficiently small values of $1/f_a$. The dark photon in that case would also not be produced appreciably (for instance, see the bottom panels of Fig.~\ref{fig:Neff:evol:subset:vert} that will be discussed later). This leads to the horizontal blue exclusion line below which the parameter space is unconstrained. 
%
%

On the other hand, for the situation with a smaller $T_\nu/T_\gamma$ than the typical value in the SM, a larger axion-to-photon temperature ratio can be more easily achievable. This happens automatically when the dark photon with mass near the BBN scale decays to the axion and photon around or after neutrino decoupling. The entropy of the dark photon is dumped mostly into the photon sector, thus decreasing the neutrino-to-photon temperature ratio. An increased $T_a/T_\gamma$ is then compensated by a smaller $T_\nu/T_\gamma$ in the overall contribution to $N_\text{eff}$. The decoupling of the axion from the thermal bath can be delayed, allowing for stronger couplings with a smaller value of $f_a$. In this situation, we expect to see a nontrivial exclusion curve near the BBN scale in the plane $(m_{\gamma'}, 1/f_a)$, as is confirmed by the island region in the right panel of Fig.~\ref{fig:temp}. The horizontal exclusion strip around roughly $1/f_a \sim 10^{-7}$ GeV$^{-1}$ that separates the two allowed regions occurs due to an intermediate regime where the dark photon is no longer sufficiently abundant to allow for a large enough modification of the neutrino temperature but still contributes an appreciable amount of axion dark radiation. For weaker couplings, at larger $f_a$, neither the dark photon nor axion are produced significantly enough to affect CMB bounds. 

\begin{figure}[tp]
\begin{center}
\includegraphics[width=0.50\textwidth]{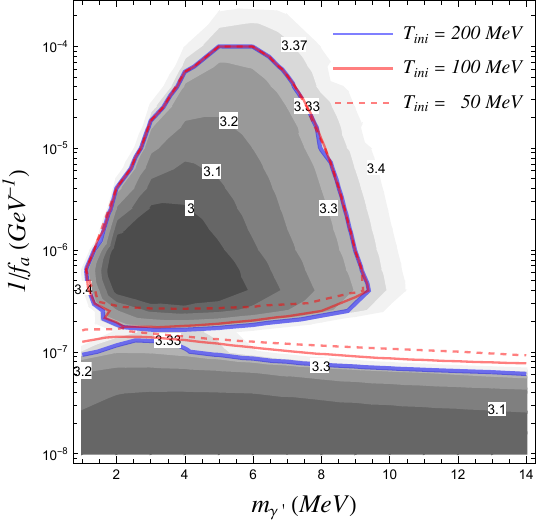}
\caption{\small Exclusion plot in the plane $(m_{\gamma'},\, 1/f_a)$ showing the contours of $N_\text{eff}$ values from simulations with different initial temperatures. The dark grey region inside the island bounded by the blue solid line and below the horizontal solid blue line with $N_\text{eff} < 3.33$ are same as Fig.~\ref{fig:temp} and they were obtained with $T_\text{ini} = 200$ MeV. Solid (dashed) red lines correspond to the allowed regions with $N_\text{eff} < 3.33$ obtained with $T_\text{ini} = 100$ (50) MeV.}
\label{fig:diff:Tini}
\end{center}
\end{figure}

We note that the size of the horizontal exclusion strip depends mildly on the assumption of the reheating temperature which determines the freeze-in abundance of dark photons in the thermal bath. In our numerical simulation of the Boltzmann equations, the initial temperature is set to be around a few hundreds of MeV as the evolution for heavier SM particles than the electron are not included. We checked that increasing the initial temperature to a higher value beyond the valid region for our numerical simulation widens slightly the horizontal exclusion band around $1/f_a \sim 10^{-7}$ GeV$^{-1}$ as more dark photons freeze-in to the thermal bath, but this widening tapers off. 
If we, instead, decrease the initial temperature down to roughly 50 MeV, the horizontal exclusion band shifts upwards slightly. As illustrated in Fig.~\ref{fig:diff:Tini}, the horizontal exclusion band moves up with decreasing initial temperature while the upper part of the unconstrained island of parameter space remains intact. This horizontal band therefore has a mild dependence on the initial reheating temperature. We discuss further the underlying physics with supplementary plots in Appendix~\ref{app:sec:supp} and Fig.~\ref{fig:Neff:evol:subset:varyingTini}.

We can see more quantitatively the aspects discussed above by following the cosmological evolution of the individual energy densities and temperature ratios. Three benchmark points with $f_a = 10^{5}$, $10^{6}$, and $10^{7}$ GeV and $m_{\gamma'} =$ 4 MeV, corresponding to the black triangle, circle and diamond points of Fig.~\ref{fig:temp}, are illustrated in Fig.~\ref{fig:Neff:evol:subset:vert} (similar plots for varying $m_{\gamma'}$ with fixed $f_a$ are illustrated in Fig.~\ref{fig:Neff:evol:subset:hori:fa6} in Appendix~\ref{app:sec:supp}). The left and right panels show the evolution of the energy densities for each species in terms of their $N_\text{eff}$ and their temperature ratio with respect to the photon temperature, respectively. In the left panels, the horizontal green band represents the range of $N_\text{eff}$ that is compatible with observation, and the solid black line is the total value of $N_\text{eff}$ that we computed. The dashed black and dot-dashed black lines represent the $N_\text{eff}$ contributions from neutrinos and axions, respectively, and the solid red line shows the evolution of the dark photon energy density. We see that the dark photon decay affects indirectly the neutrino energy density that evolves to lie outside the green band, but the dark radiation contribution compensates to allow the total $N_\text{eff}$ to remain compatible with CMB observations. In the right panels, the horizontal grey lines correspond to the expected value of the neutrino to photon temperature ratio in the SM, $T_\nu / T_\gamma = (4/11)^{1/3}$. The solid black line is the electron neutrino temperature, while the dashed and dot-dashed lines correspond to the other neutrino species' temperature and the axion temperature, respectively, all relative to the photon temperature. The enhancement in the dark radiation from the axion and the deviation of the neutrino temperature from the SM case is clearly visible. The top and middle panels confirm our picture of the physics leading to the island region. The bottom panels illustrate the situation when the dark photon freezes in at weaker couplings, which leads to the horizontal plateau in the right plot of Fig.~\ref{fig:temp}.

\begin{figure}[tp]
\begin{center}
\includegraphics[width=0.50\textwidth]{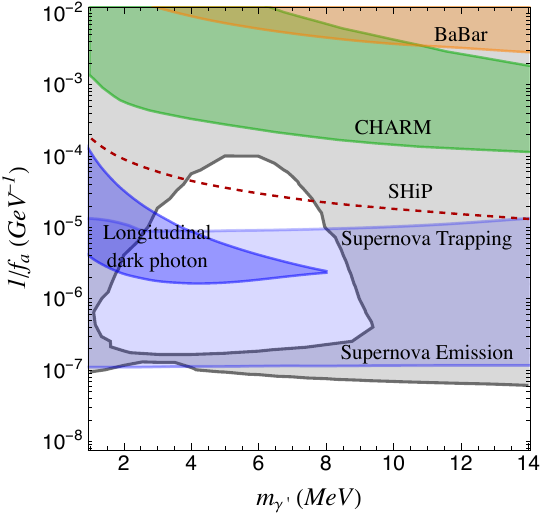}
\caption{\small Our result showing the region excluded by $\Delta N_\text{eff}$ in grey, overlaid with the constraints from supernova cooling (blue) \cite{Hook:2021ous}, CHARM (green) \cite{CHARM:1985anb}, and BaBar (orange) \cite{BaBar:2001yhh}. The projected sensitivity from SHiP \cite{SHiP:2015vad,Alekhin:2015byh} is denoted by a dashed red line.}
\label{fig:Hook}
\end{center}
\end{figure}

In Fig.~\ref{fig:Hook}, our allowed (white shaded) and excluded (grey shaded) regions in the plane $(m_{\gamma'}, 1/f_a)$ is overlaid with other constraints from supernova cooling and collider experiments as presented in Refs.~\cite{deNiverville:2018hrc,Hook:2021ous,Jodlowski:2023sbi, Jodlowski:2023yne}. The constraint on $1/f_a \sim 10^{-7}$ GeV$^{-1}$ corresponding to the horizontal plateau is stronger than the conservative BBN bound derived in Ref.~\cite{Hook:2021ous}, and reaches the lower end of the blue-shaded constraints from supernova emission of Ref.~\cite{Hook:2021ous}. The BaBar~\cite{BaBar:2001yhh} and CHARM experiments~\cite{CHARM:1985anb} are shaded in orange and green respectively. We see that they are not able to probe the allowed parameter space in the island region that lies above the supernova constraints. The proposed SHiP experiment~\cite{SHiP:2015vad,Alekhin:2015byh}, on the other hand, could have the necessary sensitivity to cover some of the parameter space, as shown by the dashed red line.

\begin{figure}[tp]
\begin{center}
\includegraphics[width=0.440\textwidth]{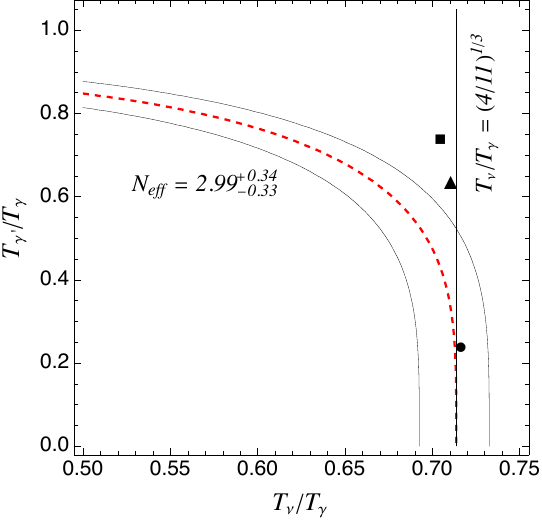}\quad
\includegraphics[width=0.45\textwidth]{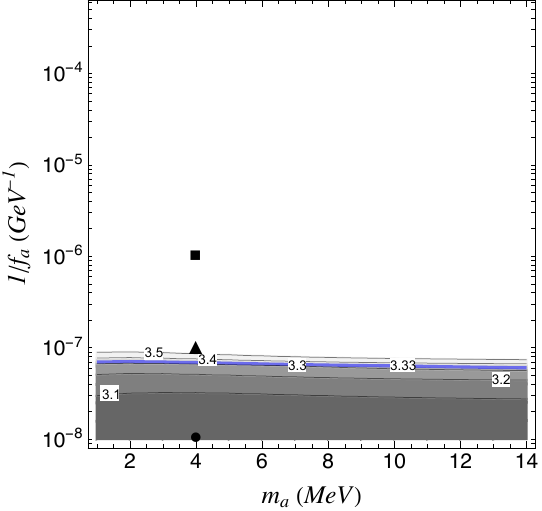}
\caption{\small Constrained regions for the scenario with a massive MeV-scale axion and lighter dark photon. Left: Allowed $N_\text{eff}$ band for the neutrino and dark photon temperatures relative to the photon temperature, assuming instantaneous decoupling. Right: Exclusion plot in the $(m_a, 1/f_a)$ plane showing contours of $N_\text{eff}$ values. The region below the blue-coloured line with $N_\text{eff} < 3.33$ is allowed.}
\label{fig:heavyaxion:lightdarkphoton}
\end{center}
\end{figure}

Finally, we show in Fig.~\ref{fig:heavyaxion:lightdarkphoton} the plots corresponding to those in Fig.~\ref{fig:temp} but for the opposite scenario of a heavy MeV-scale axion with a light dark photon, with the same axes as in Fig.~\ref{fig:temp}. Since the collision terms involved are identical, the only difference is in the number of degrees of freedom. Replacing $g_a$ with $g_{\gamma'}$ and $T_a$ with $T_{\gamma'}$ in the expression for $N_\text{eff}$ in Eq.~\ref{eq:Neff:T:darkphoton}, we see that there is a factor of two enhancement from the relativistic dark photon having two degrees of freedom instead of one as for the axion. The modified neutrino temperature is then no longer sufficient to compensate for the dark radiation contribution, as shown by the triangle and square black points on the left plot that are in the excluded white region in the right plot. In this case there is no island feature in the exclusion plane.

\section{Relaxing the cosmological bound on neutrino masses}
\label{sec:neutrino}

In the island region identified in the previous Section~\ref{sec:Neff}, the modified neutrino energy density leads to a relaxed CMB bound on neutrino masses compared to the standard scenario with no extra dark radiation. This can be seen by the expression for the neutrino energy density, $\rho_\nu = \sum m_\nu n_\nu $, when neutrinos become non-relativistic, where the sum of neutrino masses can be factorised out if the number density is universal (which is a good assumption since the difference between the muon and tau neutrino energy densities and the electron neutrino energy density is negligible according to Fig.~\ref{fig:Neff:evol:subset:vert}). Using CMB measurements, the neutrino abundance with respect to the critical density $\rho_\text{crit}^0$ is approximately given by
\begin{equation}
\Omega_\nu \equiv \frac{\rho_\nu^0}{\rho_\text{crit}^0} \simeq \frac{\sum m_\nu}{93.14 \, h^2\, \text{eV}}~,
\end{equation}
where $h$ is the reduced Hubble parameter. 
The Planck 2018 data together with BAO, polarisation, and lensing leads to a strong upper bound on the sum of neutrino masses~\cite{Planck:2018vyg},
$\sum m_\nu < 0.12 \, {\rm eV}$ at 95\% CL.
A more recent combination from Ref.~\cite{DiValentino:2021hoh} leads to a stronger upper bound,
$\sum m_\nu < 0.09 \, {\rm eV}$ at 95\% CL.
Taking the latter constraint, a reduced neutrino temperature relative to the photon temperature, $T_\nu/T_\gamma$, will then roughly relax the bound on neutrino masses by a factor of $n_\nu^\text{SM}/n_\nu$, 
\begin{equation}\label{eq:numass:relax}
\sum m_\nu < 0.09 \, {\rm eV} \,  \times \, \frac{n^{\rm SM}_\nu}{n_\nu}~,
\end{equation}
where $n_\nu \propto (T_\nu/T_\gamma)^3$ and $n^{\rm SM}_\nu$ is the neutrino number density in the SM. 
In the near future, data from the ground-based DESI (Dark Energy Spectroscopic Instrument) experiment~\cite{DESI:2016fyo} and the EUCLID satellite~\cite{Amendola:2016saw} are expected to reach a precision for the sum of neutrino masses up to $0.02 \, {\rm eV}$ at the $1\sigma$ level. 
%
\begin{figure}[t]
\begin{center}
\includegraphics[width=0.45\textwidth]{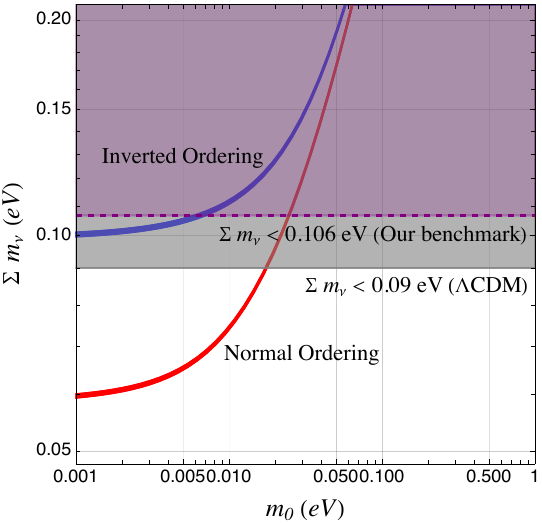}
\caption{\small Cosmological and terrestrial bounds on neutrino masses. The red (blue) colored band is the allowed sum of the neutrino masses as a function of the lightest neutrino mass $m_0$ for the normal (inverted) ordering. The gray region is excluded at $95\, \%$ CL from the most recent cosmological data~\cite{DiValentino:2021hoh}. The exclusion region is relaxed to the purple region for our benchmark point of $m_{\gamma^\prime} = 3 \, {\rm MeV}$ and $f^{-1}_a = 1.6 \times 10^{-5} \, {\rm GeV^{-1}}$ in the dark axion portal model.}
\label{fig:neutrinomass:bound}
\end{center}
\end{figure}

Neutrino oscillation experiments~\cite{deSalas:2020pgw, Esteban:2020cvm, Capozzi:2021fjo} provide a complementary probe of neutrino masses. They constrain the difference in squared neutrino masses, $\Delta m_\nu^2$, and put terrestrial constraints on $\sum m_\nu$ as a function of the lightest neutrino mass $m_0$. 
These are illustrated in Fig.~\ref{fig:neutrinomass:bound}, where the value of $\sum m_\nu$ is constrained to be inside the red and blue coloured band for the normal and inverted neutrino mass orderings respectively. We see that the sum of neutrino masses must be at least larger than $0.06 \, {\rm eV}$ ($0.10 \, {\rm eV}$) for the normal (inverted) ordering. The cosmological bounds from the CMB in the standard $\Lambda$CDM model are shown by the solid black horizontal line, which excludes the grey region above corresponding to $\sum m_\nu < 0.09$ eV~\cite{DiValentino:2021hoh}. The inverted ordering would appear to be ruled out by the CMB, which would lead to a tension in this scenario between terrestrial and cosmological observations. However, for a conservative benchmark, we illustrate how the CMB constraint can be relaxed in the presence of a dark axion portal with $m_{\gamma^\prime} = 3 \, {\rm MeV}$ and $f^{-1}_a =1.6 \times 10^{-5} \, {\rm GeV^{-1}}$. In this case the neutrino to photon temperature ratio falls down to $T_\nu / T_\gamma \simeq 0.675$. The corresponding exclusion is shown by the purple region above the horizon dashed line in Fig.~\ref{fig:neutrinomass:bound}, where the inverted ordering is now compatible with this cosmology for $m_0 \lesssim 0.005$ eV.

The mechanism pointed out here can only mildly relax the cosmological bound from the CMB. As $N_\text{eff}$ bounds get stronger due to data from DESI and EUCLID, and in the future from proposed instruments such as CMB-S4~\cite{CMB-S4:2016ple}, the constraint could be further relaxed in other approaches such as those proposed in Refs.~\cite{Farzan:2015pca,Escudero:2022gez}. Since they directly affect the neutrino sector, larger modifications of the neutrino temperature are possible. Other possibilities include using BSM neutrino decays~\cite{Chacko:2019nej,Chacko:2020hmh,Escudero:2020ped,Barenboim:2020vrr,FrancoAbellan:2021hdb} or other exotic modifications such as time-dependent neutrino masses or non-standard neutrino momentum distributions~\cite{Lorenz:2018fzb,Lorenz:2021alz,Esteban:2021ozz,Cuoco:2005qr,Oldengott:2019lke,Alvey:2021xmq}. Some variation of these scenarios may naturally be combined with the dark axion portal in a more UV-complete dark sector model. We leave the investigation of these possibilities to future studies.

\section{Cosmological constraint from BBN}
\label{sec:BBN}

\begin{figure}[tp]
\begin{center}
\includegraphics[width=0.46\textwidth]{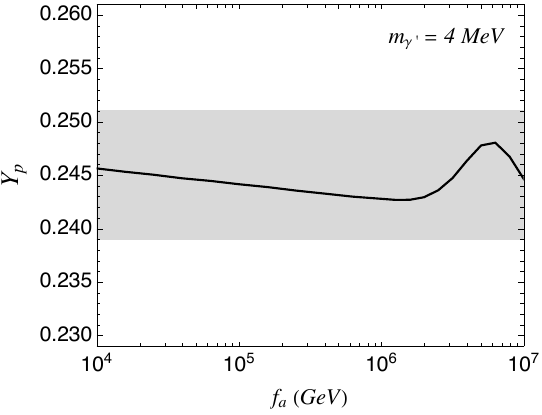}\quad
\includegraphics[width=0.46\textwidth]{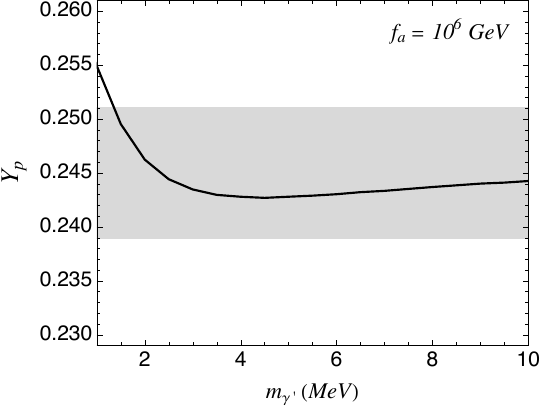}
\caption{\small The Helium abundance $Y_p$ as a function of the interaction strength $f_a$ for $m_{\gamma'} = 4$ MeV (left) and as a function of the dark photon mass $m_{\gamma'}$ for $f_a = 10^{6}$ GeV (right) in the heavy dark photon scenario. $T_D = 0.07$ MeV was chosen in both plots. The grey band corresponds to $Y_p = 0.245 \pm 0.006$ at 2$\sigma$~\cite{PhysRevD.98.030001}.}
\label{fig:BBN:TD007MeV}
\end{center}
\end{figure}

Consider first a simplified setting where only the neutrino decoupling temperature is modified, assuming the standard photon temperature. Then consistency with the observed primordial abundance of deuterium and helium requires a neutrino decoupling temperature $T_\nu^\text{dec.} > 0.6$ and $0.3$ MeV, respectively, at the $1\sigma$ level~\cite{Sabti:2019mhn}. This bound was derived by solving for the neutrino decoupling temperature with a modified Fermi constant $G_F'$ without adding any new BSM species. The modified $G_F'$ affects the evolution of all species including the primordial abundances. The neutrino temperature in the SM is $T_{\nu,\, \text{SM}}^\text{dec.} \sim1.9$ MeV, which grows with smaller $G_F'$ and reduces with larger $G_F'$, whereas the ballpark of $T_{\nu}^\text{dec.}$ in our scenario is $\mathcal{O}(\text{MeV})$ and the Fermi constant in Eq.~\ref{eq:BBN:pn} is the same as the SM. In this simplified case the primordial abundance is not expected to put any constraint on our scenario. 

\begin{figure}[tp]
\begin{center}
\includegraphics[width=0.46\textwidth]{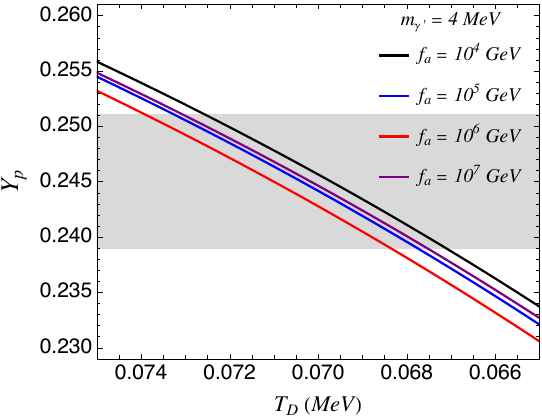}\quad
\includegraphics[width=0.46\textwidth]{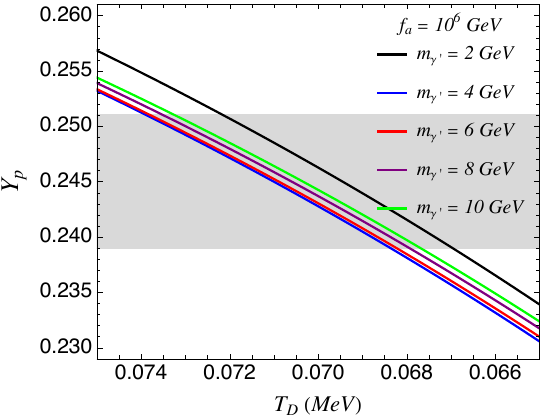}
\caption{\small The Helium abundance $Y_p$ as a function of $T_D$ for various interaction strength $f_a$ with $m_{\gamma'} = 4$ MeV (left) and various dark photon mass $m_{\gamma'}$ with $f_a = 10^{6}$ GeV (right) in the heavy dark photon scenario. The gray band corresponds to $Y_p = 0.245 \pm 0.006$ at 2$\sigma$~\cite{PhysRevD.98.030001}.}
\label{fig:BBN}
\end{center}
\end{figure}

However, the dark photon decaying into axions and photons around the BBN temperature injects some energy into the thermal bath of the visible sector, thus modifying the photon temperature. We calculate the primordial Helium abundance by using the exact evolutions of the neutrino and photon temperatures obtained from the Boltzmann equations in Eq.~\ref{eq:Boltzmann} together with the Boltzmann equations of the protons and neutrons in Eq.~\ref{eq:BBN:pn}. Following Ref.~\cite{Escudero:2018mvt} (see Appendix A.4), the Helium abundance $Y_p$ was estimated as $Y_p \simeq 2 X_n|_{T_D}$ with $T_D = 0.07$ MeV and we picked some representative benchmark points within the currently unconstrained parameter space. $T_D$ is the temperature at which Deuterium is no longer dissociated by photons~\cite{Sarkar:1995dd,Pitrou:2018cgg} and thus all neutrons at that time are expected to form Helium. The resulting primordial Helium abundance $Y_p$ from our calculation is illustrated in Fig.~\ref{fig:BBN:TD007MeV} as a function of $f_a$ for fixed $m_{\gamma'} = 4$ MeV (left panel), and varying $m_{\gamma'}$ with fixed $f_a = 10^6$ GeV (right panel). The $f_a$ value around the bump in the left panel of Fig.~\ref{fig:BBN:TD007MeV} matches to the exclusion band in Figs.~\ref{fig:temp} and~\ref{fig:Hook}. We see that the benchmark points of our parameter space in Figs.~\ref{fig:temp} and~\ref{fig:Hook} are compatible with the constraint from the Helium abundance, denoted by the horizontal grey band~\cite{PhysRevD.98.030001}. The exact value of $T_D \simeq 0.07$ MeV and the resulting primordial helium abundance will depend on a more involved numerical simulation that is beyond the scope of this work. We illustrate the sensitivity of $Y_p$ to $T_D$ in Fig.~\ref{fig:BBN} for completeness.

We do not expect the Deuterium abundance to lead to stronger constraints in our scenario as the evolution of the energy density in Fig.~\ref{fig:Neff:evol:subset:vert} and~\ref{fig:Neff:evol:subset:hori:fa6} indicates that the dark photon decay occurs around the temperature of 1 MeV and the photon injection continues until  $N_\text{eff}$ becomes saturated around $T_{\gamma} \sim 0.1$ MeV. The resulting neutrino-to-photon temperature may then affect both $N_\text{eff}$ and the Deuterium abundance in a correlated way, and a large portion of the currently unconstrained parameter space may survive when the deviation of $N_\text{eff}$ from the SM value is small. Our study motivates a more comprehensive numerical analysis for the Deuterium abundance that we leave for future work.

\section{Conclusion}
\label{sec:conclusion}

The dark axion portal is an interaction between the axion, photon and dark photon. It is motivated by non-minimality of the dark sector, where axions and dark photons have long been considered potential SM extensions in many BSM scenarios. If both are present, a dark axion portal interaction term is mandated by EFT and may significantly alter the phenomenology. 

We computed the Boltzmann equations for the evolution of the thermal bath in the early universe including the axion and dark photon interacting dominantly with the visible sector through the dark axion portal. The dark photon is taken to be at the MeV scale, with the axion much lighter. We found that the cosmological constraints from the number of relativistic degrees of freedom, $N_\text{eff}$, in CMB observations are generally stronger than astrophysical and collider bounds, and identified an island region of parameter space that is currently unconstrained. The weakening of the constraints in that region is due to the neutrino energy density being indirectly modified by the dark photon decays, with axionic dark radiation compensating to maintain an $N_\text{eff}$ value compatible with observation. This also has the effect of relaxing the cosmological bound on the sum of neutrino masses which no longer excludes the inverted ordering scenario that is still compatible with terrestrial neutrino mass constraints. 

Future experiments such as SHiP could probe a part of this open window on the dark axion portal at relatively strong coupling, or equivalently for small decay constants $f_a \sim 10^4$ GeV. Better CMB observations will be necessary to cover the entire island region parameter space and further improve the sensitivity to larger decay constants. Should a potential BSM signal emerge in cosmological $N_\text{eff}$ measurements or in terrestrial neutrino mass determinations, there may be some non-trivial interplay between the two that could give us a handle on the physics of the dark sector.

\section*{Acknowledgments}
We thank Miguel Escudero, Tae Hyun Jung, Wan-il Park, and Seokhoon Yun for useful discussions.
HH, UM, and MS were supported by National Research Foundation of Korea under Grant Number NRF-2021R1A2C1095430. TY was supported by a Branco Weiss Society in Science Fellowship and United Kingdom STFC grant ST/T000759/1.

\appendix

\section{Collision terms}
\label{app:collision}

In this section, we derive general formulae for collision terms that we used in our analysis. While, in principle, the multi-dimensional numerical integration of collision terms should be possible, the overall performance is improved if the number of integration variables are reduced before the numerical integration.
To simplify the discussion below, we distinguish two types of collision terms, denoted by $\mathcal{C}$ and $C$, contributing to the evolution of the number density and energy density, respectively. Recall that the energy density is defined as 
\begin{equation}
  \rho = \int \frac{d^3\vec{p}}{(2\pi)^3}\, g E f~,
\end{equation}
where $g$ is a degree of freedom of the particle and $f$ is the distribution function of a given species. The collision term $C$ is obtained by convoluting $\mathcal{C}$ with the energy $E$ of the corresponding species.

\subsection{Three-point collision term}
The 3-point collision term is relevant for $\gamma' \leftrightarrow a \gamma $ and the one in the evolution of the number density is given by, for the $1 \leftrightarrow 2 + 3$ process,
\begin{equation}
\begin{split}
\mathcal{C}_3 =& \int \frac{d^3\vec{p}_1}{(2\pi)^3 2 E_1}  \frac{d^3\vec{p}_2}{(2\pi)^3 2 E_2}  \frac{d^3\vec{p}_3}{(2\pi)^3 2 E_3}
 (2\pi)^4 \delta^{(4)}(p_{1\mu} - p_{2\mu} - p_{3\mu}) 
 \\[5pt]
  &\quad \times \left | \mathcal{M}_3 \right |^2 \big [ f_1 (1\pm f_2) (1\pm f_3)  - f_2 f_3 (1\pm f_1)  \big ]~.
\end{split}
\end{equation}
Assuming the Maxwell-Boltzmann distributions for the decayed particles 2 and 3 and the Bose-Einstein statistics for the particle 1, and the matrix element $\mathcal{M}$ can be factored out of integrals, the above expression simplifies to 
\begin{equation}\label{eq:col:3pt:num}
\begin{split}
\mathcal{C}_3 =& \frac{ \left | \mathcal{M}_3 \right |^2}{4 (2\pi)^3} \int_{m_1}^{\infty} dE_1 f_1 \int_{E_2^-}^{E_2^+} dE_2
\left ( 1- \exp \Big [ E_1 \left ( \frac{1}{T_1} - \frac{1}{T_3} \right ) - E_2 \left ( \frac{1}{T_2} - \frac{1}{T_3} \right )  \Big ] \right )~,
\end{split}
\end{equation}
where
\begin{equation}
  E_2^\pm = \frac{1}{2m_1^2} \left [ E_1 (m_1^2 + m_2^2 - m_3^2 ) \pm p_1
   \sqrt{\left [ (m_1 + m_2)^2 - m_3^2 \right ] \left [ (m_1 - m_2)^2 - m_3^2 \right ]}\ \right ]~.
\end{equation}
The term in Eq.~(\ref{eq:col:3pt:num}) should be convoluted with the energies $E_1$, $E_2$, and $E_3 = E_1 - E_2$ for particles 1, 2, and 3, respectively, to convert to collision terms in the Boltzmann equation for the energy density:
\begin{equation}\label{eq:col:3pt:rho}
\begin{split}
C^i_3 =& \frac{ \left | \mathcal{M}_3 \right |^2}{4 (2\pi)^3} \int_{m_1}^{\infty} dE_1 f_1 \int_{E_2^-}^{E_2^+} dE_2\, E_i
\left ( 1- \exp \Big [ E_1 \left ( \frac{1}{T_1} - \frac{1}{T_3} \right ) - E_2 \left ( \frac{1}{T_2} - \frac{1}{T_3} \right )  \Big ] \right )~.
\end{split}
\end{equation}

\subsection{Four-point collision term}

The 4-point collision term inducing the evolution of the number density for $1 + 2 \leftrightarrow 3 + 4$ is
\begin{equation}
\begin{split}
\mathcal{C}_4 =& \int \frac{d^3\vec{p}_1}{(2\pi)^3 2 E_1}  \frac{d^3\vec{p}_2}{(2\pi)^3 2 E_2}  \frac{d^3\vec{p}_3}{(2\pi)^3 2 E_3}
\frac{d^3\vec{p}_4}{(2\pi)^3 2 E_4}
 (2\pi)^4 \delta^{(4)}(p_{1\mu} + p_{2\mu} - p_{3\mu} - p_{4\mu}) 
 \\[5pt]
  &\quad \times \left | \mathcal{M}_4 \right |^2 \big [ f_1 f_2 (1\pm f_3) (1\pm f_4)  - f_3 f_4 (1\pm f_1)(1\pm f_2)  \big ]~.
\end{split}
\end{equation}
We reduce the number of integration variables for a better analytic understanding and for better performance of the numerical evaluation. While it is a difficult task for the general situation taking into account quantum statistics for all particles, as was argued in Section~\ref{sec:Boltzmann}, we assume the Maxwell-Boltzmann distributions for all particles.
\begin{equation}\label{eq:col:4pt:num:step1}
\begin{split}
\mathcal{C}_4 =& \int \frac{p_1^2 d p_1 \cdot 4\pi}{(2\pi)^3 2 E_1}  
\frac{p_2^2 d p_2 d\cos\eta \cdot 2\pi}{(2\pi)^3 2 E_2}  \frac{p_3^2 d p_3 d\cos\theta d\phi}{(2\pi)^3 2 E_3}
 (2\pi) \delta ( \left ( p_{1\mu} + p_{2\mu} - p_{3\mu} \right )^2 - m_4^2) 
 \\[5pt]
  &\quad \times \left | \mathcal{M}_4 \right |^2 \big [ f_1 f_2   - f_3 f_4  \big ]~,
\end{split}
\end{equation}
where $\eta$ is the polar angle of $\vec{p}_2$ with respect to $\vec{p}_1$ and $\theta$ and $\phi$ are polar and azimuthal angles of $\vec{p}_3$ with respect to $\vec{p}_1$. The factor $4\pi$ ($2\pi$) accounts for the freedom in choosing the orientation of $\vec{p}_1$ ($\vec{p}_2$ relative to $\vec{p}_1$). Since $\mathcal{M}_4$ is given in terms of the Mandelstam variables, we convert the integrations over $\eta$ and $\theta$ to those in terms of $s$ and $t$ whereas the integration over $\phi$ is removed using the delta function in Eq.~(\ref{eq:col:4pt:num:step1}).
\begin{equation}\label{eq:col:4pt:num:step2}
\begin{split}
\mathcal{C}_4 =& \frac{1}{8 (2\pi)^6} \int dE_1 dE_2 
\int_{s^-}^{s^+} ds \int_{t^-}^{t^+} dt \int_{E_3^-}^{E_3^+} dE_3 
 \\[5pt]
  &\quad \times  \frac{1}{ \sqrt{\left ( s - (m_1 - m_2)^2 \right ) \left ( s - (m_1 + m_2)^2 \right )}}
  \frac{\left | \mathcal{M}_4 (s,\, t) \right |^2}{\sqrt{(E_3^+ - E_3)(E_3 - E_3^-)}} \big [ f_1 f_2   - f_3 f_4  \big ] ~,
\end{split}
\end{equation}
where the upper and lower bounds for the integrations over $t$, $s$, and $E_3$ are given by, respectively,
\begin{equation}
\begin{split}
& s^{\pm} = 2E_1 E_2 +m^2_1 +m^2_2 \pm 2 p_1 p_2~,
\\ & t^{\pm} = \frac{1}{2s} \left[ -(m^2_1-m^2_2) (m^2_3-m^2_4) + (m^2_1+m^2_2+m^2_3+m^2_4) s - s^2 \right.
\\ &\quad \quad \quad \left. \pm \sqrt{\left \{ s- (m_1 + m_2)^2 \right \} \left \{ s- (m_1 - m_2)^2 \right \} \left \{ s- (m_3 + m_4)^2 \right \} \left \{ s- (m_3 - m_4)^2 \right \} }\right]~,
\\ & E^\pm_3 = \left[ \left \{ s- (m_1 + m_2)^2 \right \} \left \{ s- (m_1 - m_2)^2 \right \} \right]^{-1}
\\ & \quad \quad\times  \left[ E_1\left \{ (s-m^2_1 -m^2_2) (s+t-m^2_4) -m^2_1 (s-m^2_1+m^2_2) - 2m^2_2 (m^2_3-t)\right \} \right.
\\ & \quad \quad \quad +E_2 \left \{ (s-m^2_2)(m^2_3-t)+m^2_1 (m^2_1-m^2_2-m^2_3+2m^2_4-s-t)\right \}
\\ & \quad \quad \quad \left. \pm \sqrt{s(s-s^+)(s-s^-)(t-t^+)(t-t^-)} \right]~.
\end{split}
\end{equation}

The collision term for the energy density of $i$th particle is given by 
\begin{equation}\label{eq:col:4pt:rho}
\begin{split}
C^i_4 =& \frac{1}{8 (2\pi)^6} \int dE_1 dE_2 
\int_{s^-}^{s^+} ds \int_{t^-}^{t^+} dt \int_{E_3^-}^{E_3^+} dE_3 
 \\[5pt]
  &\quad \times  \frac{E_i}{ \sqrt{\left ( s - (m_1 - m_2)^2 \right ) \left ( s - (m_1 + m_2)^2 \right )}}
  \frac{ \left | \mathcal{M}_4 (s,\, t) \right |^2}{\sqrt{(E_3^+ - E_3)(E_3 - E_3^-)}} \big [ f_1 f_2   - f_3 f_4  \big ] ~.
\end{split}
\end{equation}
In the evaluation of $C^1_4$ and $C^2_4$, the integration over $E_3$ can be removed by noting that
\begin{equation}
  \int_{E_3^-}^{E_3^+}   \frac{dE_3}{\sqrt{(E_3^+ - E_3)(E_3 - E_3^-)}} = \pi~.
\end{equation}

\section{Collision terms for the massive dark photon and light axion case}
We present here explicit forms of collision terms for the scenario with an MeV-scale axion and light photon, $m_{\gamma'} \gg m_a$ . 

\subsection{Dark photon decay}
The collision term in the evolution of the energy density for $\gamma' \leftrightarrow a \gamma$ ($1\leftrightarrow 2+3$) is copied from the formula in Eq.~(\ref{eq:col:3pt:rho}):
\begin{equation}
\begin{split}
C^i_{\gamma' \leftrightarrow a\gamma} = \frac{ \left | \mathcal{M}_3 \right |^2}{4 (2\pi)^3} \int_{m_{\gamma'}}^{\infty} dE_{\gamma'} f_{\gamma'} \int_{E_a^-}^{E_a^+} dE_a E_i
\left ( 1- \exp \Big [ E_{\gamma'} \left ( \frac{1}{T_{\gamma'}} - \frac{1}{T_\gamma} \right ) - E_a \left ( \frac{1}{T_a} - \frac{1}{T_\gamma} \right )  \Big ] \right )~,
\end{split}
\end{equation}
where the amplitude squared (summed over polarizations of initial and final states) and the upper/lower bound of the integration over $E_a$ are given by
\begin{equation}
\begin{split}
 \left | \mathcal{M}_3 \right |^2 = \frac{\left ( m_{\gamma'}^2 - m_a^2 \right )^2}{2 f_a^2}~,\quad
 E_a^\pm = \frac{1}{2m^2_{\gamma'}} \left [ E_{\gamma'} (m_{\gamma'}^2 + m_a^2) \pm p_{\gamma'} (m^2_{\gamma'} - m^2_a) \right ]~.
\end{split}
\end{equation}
For each collision term, the integration over $E_a$ can be done analytically. We obtain the following collision terms for the energy density,
\begin{equation}
\begin{split}
   C^{\gamma'}_{\gamma' \leftrightarrow a\gamma}  &= \int_{m_{\gamma'}}^\infty d E_{\gamma'} \frac{1}{4 (2\pi)^3 }
   \frac{\left ( m_{\gamma'}^2 - m_a^2 \right )^2}{2 f_a^2} \frac{E_{\gamma'}}{e^{E_{\gamma'}/T_{\gamma'}} - 1}
   \\[5pt]
   & \quad \times \left [ \frac{p_{\gamma'} (m_{\gamma'}^2 - m_a^2)}{m_{\gamma'}^2} + \frac{e^{E_{\gamma'} \left ( \frac{1}{T_{\gamma'}} - \frac{1}{T_\gamma} \right )} }{\frac{1}{T_a} - \frac{1}{T_\gamma}} 
   \left \{e^{-E_a^+ \left (\frac{1}{T_a} - \frac{1}{T_\gamma} \right )} - e^{-E_a^- \left (\frac{1}{T_a} - \frac{1}{T_\gamma} \right )}  \right \}
   \right ]~,
   \\[5pt]
   C^{a}_{\gamma' \leftrightarrow a\gamma}  &= \int_{m_{\gamma'}}^\infty d E_{\gamma'} \frac{1}{4 (2\pi)^3 }
   \frac{\left ( m_{\gamma'}^2 - m_a^2 \right )^2}{2 f_a^2} \frac{1}{e^{E_{\gamma'}/T_{\gamma'}} - 1}
   \left [ \frac{p_{\gamma'} E_{\gamma'} (m_{\gamma'}^4 - m_a^4)}{2 m_{\gamma'}^4} \right .
   \\[5pt]
   & \left . \quad  + \frac{e^{E_{\gamma'} \left ( \frac{1}{T_{\gamma'}} - \frac{1}{T_\gamma} \right )} }{\frac{1}{T_a} - \frac{1}{T_\gamma}} 
   \left \{ \left ( E_a^+ + \frac{1}{\frac{1}{T_a} - \frac{1}{T_\gamma}} \right ) e^{-E_a^+ \left (\frac{1}{T_a} - \frac{1}{T_\gamma} \right )} 
   - \left ( E_a^- + \frac{1}{\frac{1}{T_a} - \frac{1}{T_\gamma}} \right ) e^{-E_a^- \left (\frac{1}{T_a} - \frac{1}{T_\gamma} \right )}  \right \}
   \right ]~,
   \\[5pt]
   C^\gamma_{\gamma' \leftrightarrow a\gamma}  &= C^{\gamma'}_{\gamma' \leftrightarrow a\gamma}- C^a_{\gamma' \leftrightarrow a\gamma}~.
\end{split}
\end{equation} 
Note that these collision terms assume the Maxwell-Boltzmann distributions for $a$ and $\gamma$ and take into account the quantum statistics for the dark photon.

\subsection{$s$-channel process}
The amplitude squared summed over spins and polarizations for the electron pair creation and annihilation process $\gamma' a \leftrightarrow e^+e^-$ ($1+2 \leftrightarrow 3+4$) mediated by the photon is given by 
\begin{equation}
\begin{split}
 \left | \mathcal{M}_4 (s,\, t) \right |^2 &=
 \frac{4\pi \alpha}{f_a^2 s^2} \Big [ s^3 + \left ( m_{\gamma'}^4 + m_a^4 \right ) \left (s+ 2m_e^2 \right ) - 2s \left ( s+t+m_e^2 \right ) \left (m_{\gamma'}^2 + m_a^2 \right )
 \\[5pt]
 & \quad + 2 st \left ( s+t \right ) + 2 m_e^2 \left ( m_e^2 s - 2st - 2 m_{\gamma'}^2 m_a^2 \right ) \Big ]~.
\end{split}
\end{equation}
Assuming the Maxwell-Boltzmann distributions for all particles, we evaluate the collision term for the energy density using the formula in Eq.~(\ref{eq:col:4pt:rho}).
The term $f_3 f_4$, or $f_{e^+} f_{e^-}$, in Eq.~(\ref{eq:col:4pt:rho}) can be expressed in terms of $E_{\gamma'}$ and $E_a$,
\begin{equation}
   f_{e^+} f_{e^-} = e^{-\frac{E_{e^-} + E_{e^+}}{T_{\gamma e}}} = e^{-\frac{E_{\gamma'} + E_{a}}{T_{\gamma e}}}~.
\end{equation}
The four-point collision terms for the dark photon and axion, $C^{\gamma'}_{\gamma' a \leftrightarrow e^+e^-}$ and $C^{a}_{\gamma' a \leftrightarrow e^+e^-}$, respectively, have the same form except the convoluted energy in the integrand for the energy density, and the one for the electron pair is obtained using energy conservation, $E_{\gamma'} + E_a = E_{e^+} + E_{e^-}$.  They are given by
\begin{equation}
\begin{split}
C^{\gamma',\, a}_{\gamma' a \leftrightarrow e^+e^-} &= \frac{1}{8 (2\pi)^6}\frac{8\pi^2 \alpha}{3 f_a^2} \int_{m_{\gamma'}}^\infty dE_{\gamma'} \int_{m_a}^\infty dE_a E_{\gamma',\, a} \left (  e^{-\frac{E_{\gamma'}}{T_{\gamma'}} - \frac{E_a}{T_a}} - e^{-\frac{E_{\gamma'} + E_a}{T_{\gamma}} } \right )
 \\[5pt]
 & \times \left [ \left \{ - \frac{(m_{\gamma'}^2-m_a^2)^2}{3} \left ( 5+ \frac{4m_e^2}{s^+} \right ) + 8 m_e^2 (m_{\gamma'}^2 + m_a^2 )
 \right . \right .
 \\[5pt]
& \left . \hspace{4cm} + \frac{s^+}{2} \left ( s^+ - 4 m_{\gamma'}^2 - 4 m_a^2 + 2m_e^2 \right ) \right \} \sqrt{1-\frac{4 m_e^2}{s^+}}
 \\[5pt]
 & \hspace{2.5mm} -\left \{ - \frac{(m_{\gamma'}^2-m_a^2)^2}{3} \left ( 5+ \frac{4m_e^2}{s^-} \right ) + 8 m_e^2 (m_{\gamma'}^2 + m_a^2 )
 \right .
 \\[5pt]
 & \left . \hspace{4cm} + \frac{s^-}{2} \left ( s^- - 4 m_{\gamma'}^2 - 4 m_a^2 + 2m_e^2 \right ) \right \} \sqrt{1-\frac{4 m_e^2}{s^-}}
  \\[5pt]
 & \left . \hspace{2.5mm} + 2 \left ( \left ( m_{\gamma'}^2 - m_a^2 \right )^2 - 6 m_e^4 \right ) \log \frac{\sqrt{s^+} + \sqrt{s^+ - 4m_e^2}}{\sqrt{s^-} + \sqrt{s^- - 4m_e^2}}\  \right ]~,
 \\[5pt]
 C^{e}_{\gamma' a \leftrightarrow e^+e^-}  &= C^{\gamma'}_{\gamma' a \leftrightarrow e^+e^-}  + C^{a}_{\gamma' a \leftrightarrow e^+e^-}~,
\end{split}
\end{equation}
where the upper and lower bound of $s$ are given by
\begin{equation}
  s^\pm = 2 E_{\gamma'} E_a + m_{\gamma'}^2 + m_a^2 \pm 2 p_{\gamma'} p_a~.
\end{equation}

\subsection{$t$-channel process}
The squared matrix element for $t$-channel process $e^{\pm} \gamma^\prime \leftrightarrow e^{\pm} a$ is given by
\begin{equation}
\begin{split}
 & \left | \mathcal{M}_4 (s,\, t) \right |^2 = 
 - \frac{4\pi \alpha}{f_a^2 t^2} \Big [ t^3 + 2 \left ( s - m_{\gamma'}^2 - m_a^2 \right ) t^2 
 \\[5pt]
 &\hspace{1.5cm} + \left \{ m_{\gamma'}^4 + m_a^4 + 2 \left ( s - m_e^2 \right )^2 - 2 \left ( m_{\gamma'}^2 + m_a^2 \right ) \left ( s + m_e^2 \right ) \right \} t
 + 2 m_e^2 \left ( m_{\gamma'}^2 - m_a^2 \right )^2 \Big ]~.
\end{split}
\end{equation}
While the simplified analytic expression of the collision term is not available, it can be approximated for the massless axion ($m_a = 0$) in the limit of the massless electron, or $m_e \rightarrow 0$. The residual dependence on $m_e$ in logarithmic terms is to regulate the infrared divergence. They are given by

\begin{equation}
\begin{split}
C^{\gamma '}_{e^\pm  \gamma ' \leftrightarrow e^\pm a} &\simeq \frac{1}{8 (2\pi)^6} \frac{4\pi^2 \alpha_e}{f^2_a} \int^\infty_{m_{\gamma '}} dE_{\gamma '} \int^\infty_0 dE_e \, E_{\gamma '} e^{ -\frac{E_{\gamma '}}{T_{\gamma '}}-\frac{\strut E_e}{T_{\gamma}}} 
\\[5pt]
& \times \Big [ -4p_{\gamma '} E_e \left(13 E_{\gamma '} E_e +7m^2_{\gamma '}\right) 
\\[5pt]
& \hspace{2.5mm} +4p_{\gamma '} E_e \left(2E_{\gamma '}E_e+m^2_{\gamma '}\right) \log \frac{16E^4_e \left(4E_e \left(E_{\gamma '}+E_e\right)+m^2_{\gamma '}\right)}{m^4_e m^2_{\gamma '}}
\\[5pt] 
& \hspace{2.5mm} + \left \{ 8E^2_{\gamma '} E^2_e +4 \left( E_{\gamma '} - E_e \right) E_e m^2_{\gamma '} + m^4_{\gamma '}\right \} \log \frac{\left( E_{\gamma '}+p_{\gamma '} \right)^4 \left( 2E_{\gamma '} E_e +m^2_{\gamma '}+2p_{\gamma '} E_e \right)^2}{m^6_{\gamma '}\left(4 \left( E_{\gamma '}+E_e\right) E_e+m^2_{\gamma '}\right)}
\\[5pt] 
& \hspace{2.5mm} + 4m^4_{\gamma '}\left( \log \frac{2E_e}{m_e} -1 \right) \log \frac{E_{\gamma '}+p_{\gamma '}}{m_{\gamma '}} 
\\[5pt] 
& \left . \hspace{2.5mm} + m^4_{\gamma '} \left \{  {\rm Li}_2 \left[ -\frac{2E_e}{E_{\gamma '}+p_{\gamma '}}\right]- {\rm Li}_2 \left[ -\frac{2E_e}{m^2_{\gamma '}} (E_{\gamma '} +p_{\gamma '}) \right] \right \} \right ] 
\\[5pt] 
 & -\frac{1}{8 (2\pi)^6} \frac{4\pi^2 \alpha_e}{f^2_a} \int^\infty_0 dE_a \int^\infty_{\frac{m^2_{\gamma '}}{4E_a}} dE_e \,e^{ -\frac{E_a}{T_a}-\frac{E_e}{T_{\gamma}} }
\\[5pt]
& \times \left[ -\frac{4E_a E_e-m^2_{\gamma '}}{24 E_a E_e} \left \{ 8E^2_a E^2_e \left(47E_a-8E_e \right) \right . \right .
\\[5pt]
& \left. \hspace{4cm} +10E_a E_e \left(-17E_a +26E_e \right) m^2_{\gamma '}+\left( -2E_a +5E_e \right) m^4_{\gamma '} \right \}
\\[5pt]
& \hspace{2.5mm} +\frac{4E_a E_e -m^2_{\gamma '}}{2E_a} \left \{16 E^3_a E_e -4 E_a (E_a-2E_e) m^2_{\gamma '} +m^4_{\gamma '} \right \} \log \frac{4E_a E_e -m^2_{\gamma '}}{m_{\gamma '} m_e}
\\[5pt]
& \hspace{2.5mm} - \left \{ \frac{m^6_{\gamma '}}{4E_a}+\frac{m^4_{\gamma '}}{2} (5E_a - 8E_e) +8E_a E_e (E_a-E_e) m^2_{\gamma '} -16 E^3_a E^2_e \right \} \log \frac{4E_a E_e}{m^2_{\gamma '}}
\\[5pt]
& \hspace{2.5mm} +\frac{1}{2} (E_a-2E_e) m^4_{\gamma '}  \log \frac{4E_a E_e}{m^2_{\gamma '}} \log \frac{4E_a E_e (4E_a E_e -m^2_{\gamma '})^4}{m^6_{\gamma '} m^4_e}
\\[5pt]
& \left .  \hspace{2.5mm} +2 (E_a-2E_e) m^4_{\gamma '} {\rm Li}_2 \left[ -\frac{4E_a E_e-m^2_{\gamma '}}{m^2_{\gamma '}} \right] \right]~,
\end{split}
\end{equation}

\begin{equation}
\begin{split}
C^{a}_{e^\pm  \gamma ' \leftrightarrow e^\pm a} & \simeq \frac{1}{8 (2\pi)^6} \frac{4\pi^2 \alpha_e}{f^2_a} \int^\infty_{m_{\gamma '}} dE_{\gamma '} \int^\infty_0 dE_e \, e^{-\frac{E_{\gamma '}}{T_{\gamma '}}-\frac{\strut E_e}{T_{\gamma}}}
\\[5pt]
& \times \left[ -\frac{2}{3} p_{\gamma '} E_e \left \{ 2E_{\gamma '} E_e (47 E_{\gamma '}-8E_e) +(54E_{\gamma '}-85E_e) m^2_{\gamma '} \right \}
\right .
\\[5pt]
& \hspace{2.5mm} +4E_e \left \{ 8E^3_{\gamma '} E_e +4E_{\gamma '} (E_{\gamma '}-2E_e) m^2_{\gamma '}+3m^4_{\gamma '} \right \} \log \frac{E_{\gamma '}+p_{\gamma '}}{m_{\gamma '}}
\\[5pt]
& \hspace{2.5mm} +\frac{1}{2} \left \{ 16 E^3_{\gamma '} E^2_e +8E_{\gamma '} E_e (E_{\gamma '}-2E_e) m^2_{\gamma '} \right . 
\\[5pt]
& \left . \hspace{4cm} +(3E_{\gamma '}-2E_e) m^4_{\gamma '} \right \} \log \frac{(2E_{\gamma '} E_e +m^2_{\gamma '}+2p_{\gamma '} E_e)^2}{m^2_{\gamma '} (4E_e (E_{\gamma '}+E_e) +m^2_{\gamma '})}
\\[5pt]
& \hspace{2.5mm} +\frac{1}{2} p_{\gamma '} \left \{ 16 E^2_{\gamma '} E^2_e +8 (E_{\gamma '}-E_e) E_e m^2_{\gamma '} -m^4_{\gamma '} \right \} \log \frac{16 E^4_e (4E_e (E_{\gamma '}+E_e) +m^2_{\gamma '})}{m^2_{\gamma '} m^4_e}
\\[5pt]
&  \hspace{2.5mm} + 4(E_{\gamma '}-2E_e)m^4_{\gamma '} \log \frac{E_{\gamma '} +p_{\gamma '}}{m_{\gamma '}} \log \frac{2E_e}{m_e}
\\[5pt]
& \left . \hspace{2.5mm} + (E_{\gamma '}-2E_e)m^4_{\gamma '} \left \{ {\rm Li}_2 \left[ -\frac{2E_e}{E_{\gamma '}+p_{\gamma '}} \right] - {\rm Li}_2 \left[ -\frac{2E_e}{m^2_{\gamma '}} (E_{\gamma '}+p_{\gamma '}) \right]  \right \} \right]
\\[5pt]
& - \frac{1}{8 (2\pi)^6} \frac{4\pi^2 \alpha_e}{f^2_a} \int^\infty_0 dE_a \int^\infty_{\frac{m^2_{\gamma '}}{4E_a}} dE_e \, E_a e^{ -\frac{E_a}{T_a}-\frac{E_e}{T_{\gamma}}}
\\[5pt]
& \times \left[ -\frac{1}{4} \left(52E_a E_e -23 m^2_{\gamma '}\right) \left(4E_a E_e -m^2_{\gamma '}\right) 
\right .
\\[5pt]
& \hspace{2.5mm} +\frac{1}{2} \left( 32 E^2_a E^2_e -16E_a E_e m^2_{\gamma '} -3 m^4_{\gamma '} \right) \log \frac{4E_a E_e}{m^2_{\gamma '}}
\\[5pt]
& \hspace{2.5mm} +2 \left(4E_a E_e -m^2_{\gamma '}\right)^2 \log \frac{4E_a E_e -m^2_{\gamma '}}{m_{\gamma '} m_e } 
\\[5pt]
& \hspace{2.5mm} + \frac{1}{2}m^4_{\gamma '} \log \frac{4E_a E_e}{m^2_{\gamma '}} \log \frac{4E_a E_e (4E_a E_e -m^2_{\gamma '})^4}{m^6_{\gamma '} m^4_e} 
\\[5pt]
& \left . \hspace{2.5mm} + 2m^4_{\gamma '} {\rm Li}_2 \left[ -\frac{4E_a E_e -m^2_{\gamma '}} {m^2_{\gamma '}} \right] \right]~,
\end{split}
\end{equation}

\begin{equation}
\begin{split}
 C^{e}_{e^\pm  \gamma ' \leftrightarrow e^\pm a}  &= C^{\gamma'}_{e^\pm  \gamma ' \leftrightarrow e^\pm a} - C^{a}_{e^\pm  \gamma ' \leftrightarrow e^\pm a}~.
\end{split}
\end{equation}

\section{Supplementary plots}
\label{app:sec:supp}

\begin{figure}[tph]
\begin{center}
\includegraphics[width=0.38\textwidth]{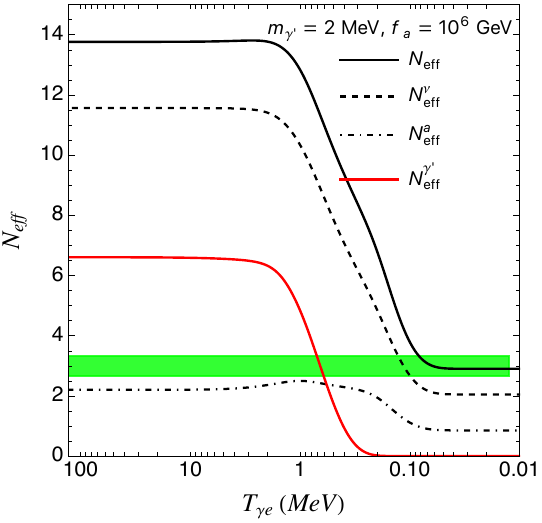}
\includegraphics[width=0.385\textwidth]{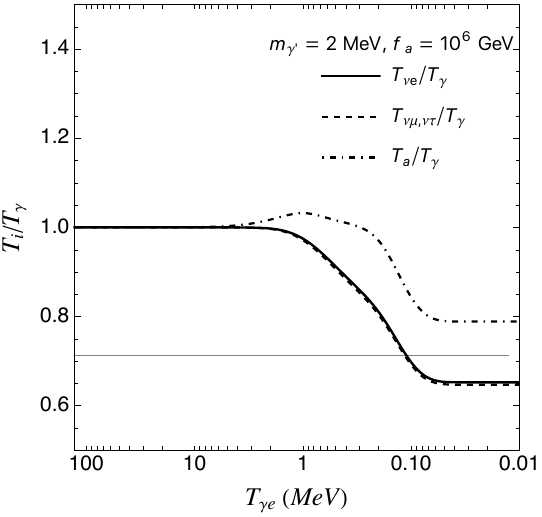}\\
\includegraphics[width=0.38\textwidth]{plots/Neff4MeVfa6.pdf}
\includegraphics[width=0.385\textwidth]{plots/TiTgam04MeVfa6.pdf}\\
\includegraphics[width=0.38\textwidth]{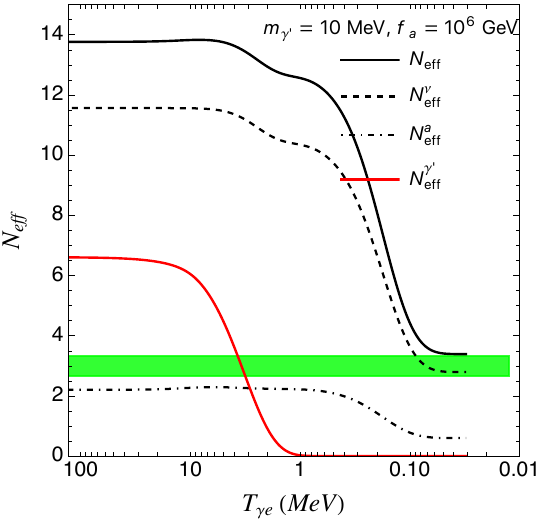}
\includegraphics[width=0.385\textwidth]{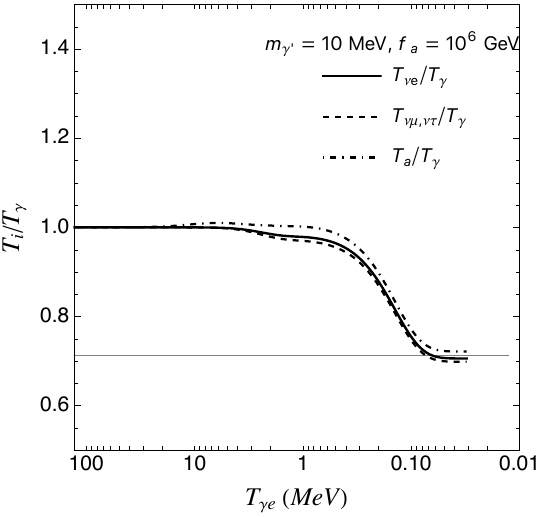}
\caption{\small Left:  evolutions of $N_\text{eff} \equiv N_\text{eff}^\nu + N_\text{eff}^a$ and individual contributions for $m_{\gamma'} = 2, 4, 10$ MeV and $f_a = 10^{6}$ GeV in heavy dark photon and light axion scenario. The green band corresponds to the allowed region, $N_\text{eff} = 2.99^{+0.34}_{-0.33}$ from Planck 2018~\cite{Planck:2018vyg}. Right: the evolution of the $i$-species-to-photon temperature ratios for the same benchmark scenarios. The gray horizontal line corresponds to $T_\nu/T_\gamma = (4/11)^{1/3}$.}
\label{fig:Neff:evol:subset:hori:fa6}
\end{center}
\end{figure}

\begin{figure}[tph]
\begin{center}
\includegraphics[width=0.38\textwidth]{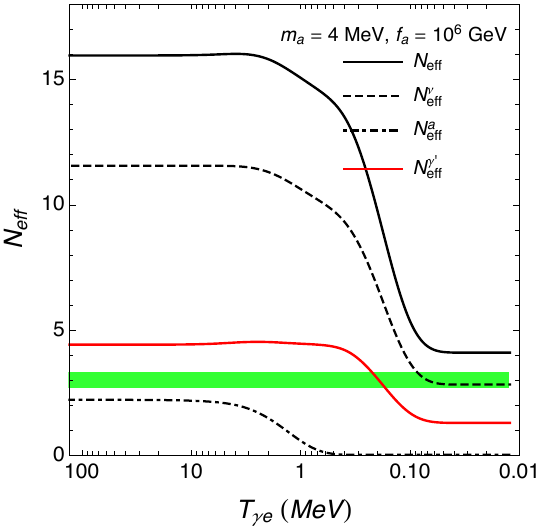}
\includegraphics[width=0.385\textwidth]{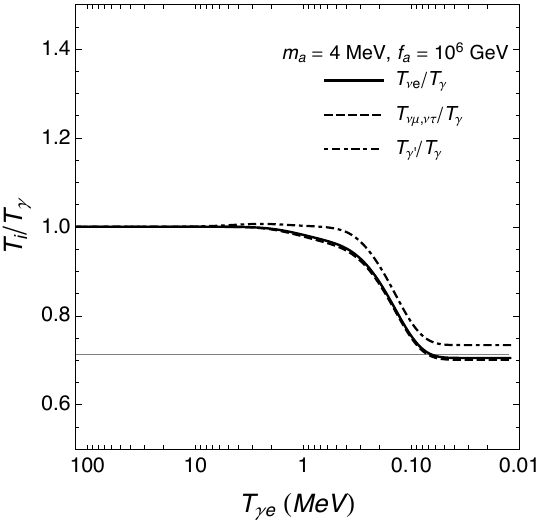}\\
\includegraphics[width=0.38\textwidth]{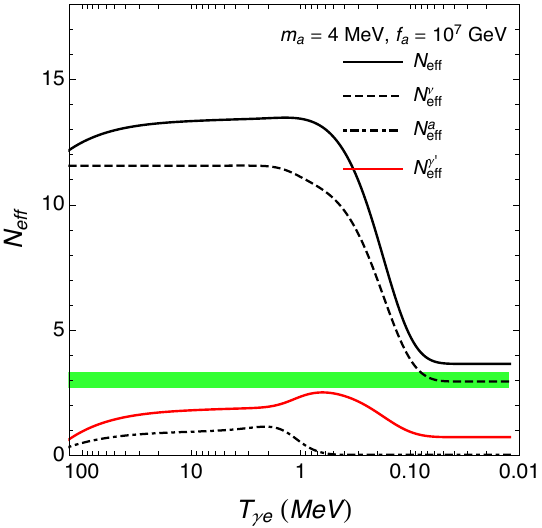}
\includegraphics[width=0.385\textwidth]{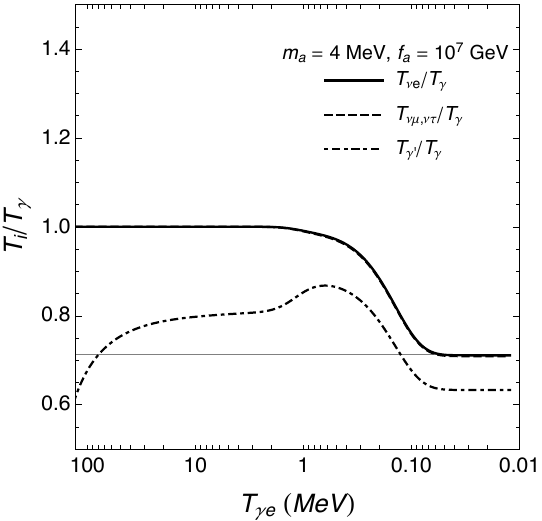}\\
\includegraphics[width=0.38\textwidth]{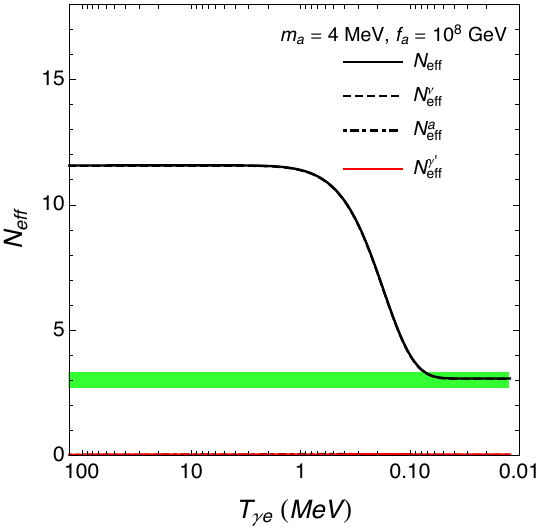}
\includegraphics[width=0.385\textwidth]{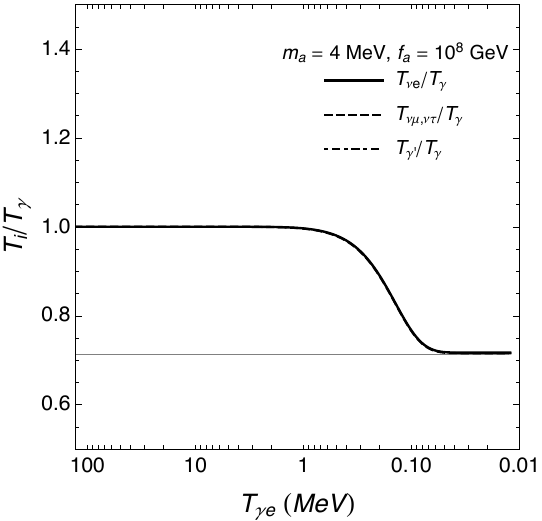}\\
\caption{\small Left:  evolutions of $N_\text{eff} \equiv N_\text{eff}^\nu + N_\text{eff}^{\gamma'}$ and individual contributions for $m_{a} = 4$ MeV and $f_a = 10^{6}, 10^{7}, 10^{8}$ GeV in the heavy axion and light dark photon scenario. The green band corresponds to the allowed region, $N_\text{eff} = 2.99^{+0.34}_{-0.33}$ from Planck 2018~\cite{Planck:2018vyg}. Right: the evolution of the $i$-species-to-photon temperature ratios for the same benchmark scenarios. The gray horizontal line corresponds to $T_\nu/T_\gamma = (4/11)^{1/3}$. In the bottom panels, curves for the dark photon and axion are not visible due to their tiny contributions.}
\label{fig:Neff:evol:subset:hori:ma4}
\end{center}
\end{figure}

\begin{figure}[tph]
\begin{center}
\includegraphics[width=0.38\textwidth]{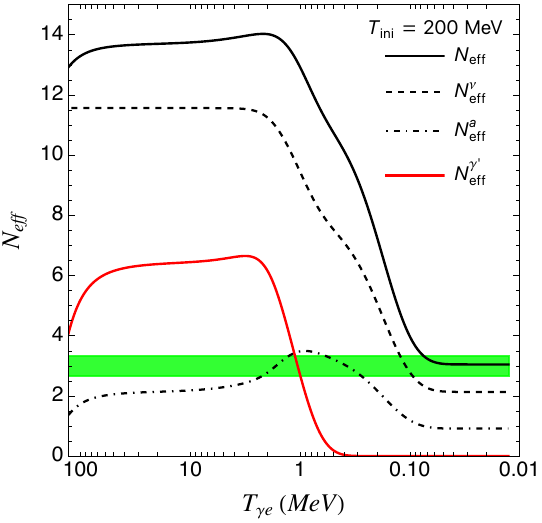}
\includegraphics[width=0.385\textwidth]{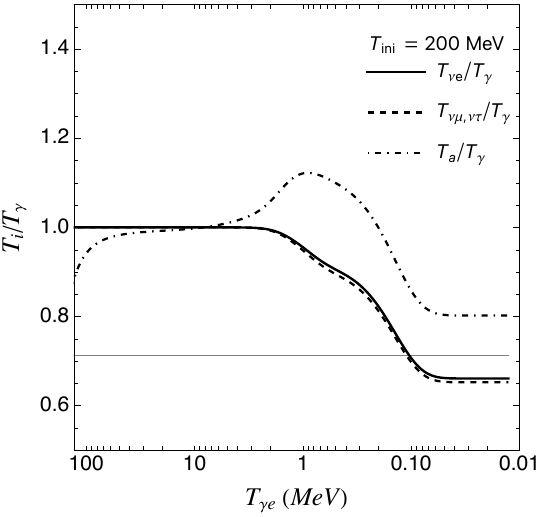}\\
\includegraphics[width=0.38\textwidth]{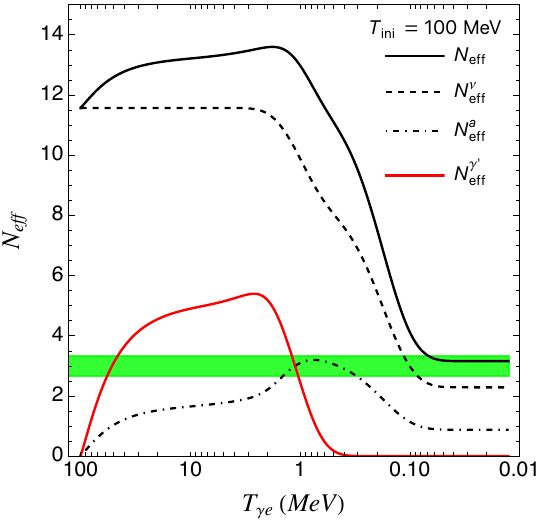}
\includegraphics[width=0.385\textwidth]{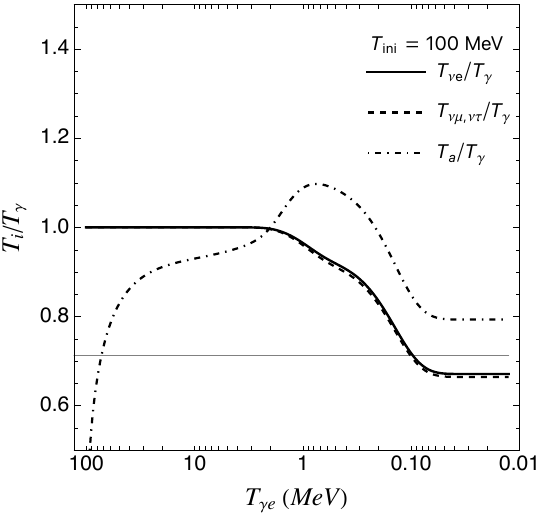}\\
\includegraphics[width=0.38\textwidth]{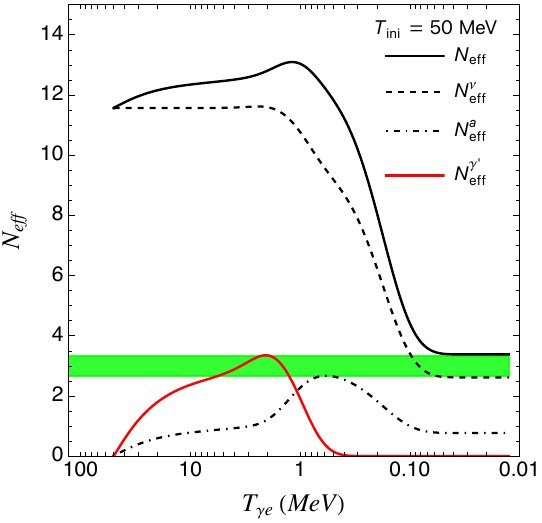}
\includegraphics[width=0.385\textwidth]{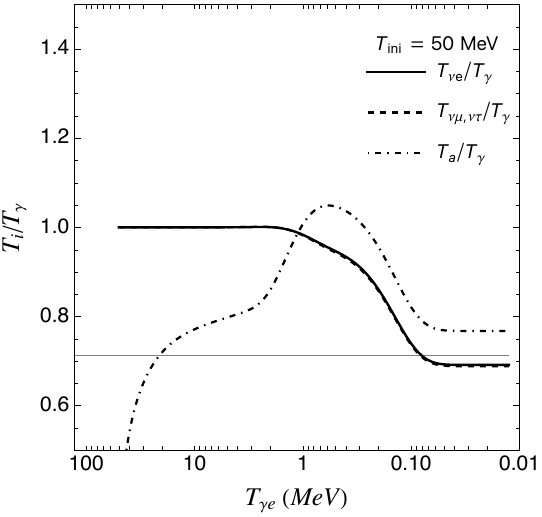}\\
\caption{\small Left:  evolutions of $N_\text{eff} \equiv N_\text{eff}^\nu + N_\text{eff}^a$ and individual contributions for $m_{\gamma'} = 4$ MeV and $f_a^{-1} = 2.51 \times 10^{-7}$ GeV$^{-1}$ in heavy dark photon and light axion scenario. The green band corresponds to the allowed region, $N_\text{eff} = 2.99^{+0.34}_{-0.33}$ from Planck 2018~\cite{Planck:2018vyg}. Right: the evolution of the $i$-species-to-photon temperature ratios for the same benchmark scenarios. The gray horizontal line corresponds to $T_\nu/T_\gamma = (4/11)^{1/3}$.}
\label{fig:Neff:evol:subset:varyingTini}
\end{center}
\end{figure}

The cosmological evolution of energy densities and temperatures for different species in the heavy dark photon and light axion scenario are provided in Fig.~\ref{fig:Neff:evol:subset:hori:fa6} for the benchmark points with the dark photon mass of $m_{\gamma'} = 2,\, 4,\, 10$ MeV and $f_a = 10^6$ GeV. Similar plots for the heavy axion and light dark photon scenario are provided in Fig.~\ref{fig:Neff:evol:subset:hori:ma4} for benchmark points with $m_a = 4$ MeV and $f_a = 10^6,\, 10^7,\, 10^8$ GeV.

In Fig.~\ref{fig:Neff:evol:subset:varyingTini}, the dependence of the cosmological evolution of energy densities and temperatures for different species on the initial temperature in the heavy dark photon and light axion scenario is shown for the benchmark point with $m_{\gamma'} = 4$ MeV and $f_a^{-1} = 2.51 \times 10^{-7}$ GeV$^{-1}$. The corresponding benchmark point is excluded for a lower reheating temperature such as $T_\text{ini} = 50$ MeV while it survives for higher reheating temperatures such as $T_\text{ini} = 100,\ 200$ MeV as was illustrated in Fig.~\ref{fig:diff:Tini}. When the reheating temperature is as low as $T_\text{ini} = 50$ MeV, the dark photon is not produced large enough to significantly modify the neutrino-to-photon temperature ratio, compared to the typical value (see the bottom plots of Fig.~\ref{fig:Neff:evol:subset:varyingTini}), and thus the previously allowed selected benchmark point get ruled out.

\newpage

{\small
\bibliography{main}{}
\bibliographystyle{JHEP}}

\end{document}